\documentclass{aa} 
\input psfig.tex
\usepackage{graphicx}
\usepackage{natbib}
\usepackage{array}
\usepackage{graphics}
\usepackage{latexsym}
\usepackage{amssymb}
\usepackage{amsmath}
\usepackage{fancyhdr}
\usepackage{morefloats}
\usepackage{bm}
\usepackage{graphicx}
\usepackage{txfonts}

\begin{document}
\title{Thin disk kinematics from RAVE and the solar motion}

\author{S. Pasetto\inst{1,2},
E.K. Grebel\inst{2}, 
T. Zwitter\inst{3,4}, 
C. Chiosi\inst{5}, 
G. Bertelli\inst{6},
O. Bienayme\inst{7},
G. Seabroke\inst{1},
J. Bland-Hawthorn\inst{8},
C. Boeche\inst{2}, 
B.K. Gibson\inst{9,10},
G. Gilmore\inst{11},
U. Munari\inst{6}, 
J.F. Navarro\inst{12},
Q. Parker\inst{13,14,15},
W. Reid\inst{13,14},
A. Silviero\inst{5,16},
M. Steinmetz\inst{16}}

\offprints{sp2@mssl.ucl.ac.uk}

\institute{ 
University College London, Department of Space \& Climate Physics, Mullard Space Science Laboratory, Holmbury St. Mary, Dorking Surrey RH5 6NT, United Kingdom
\and 
Astronomisches Rechen-Institut, Zentrum f\"ur Astronomie der Universit\"at Heidelberg, M\"onchhofstr. 12-14, 69120 Heidelberg, Germany
\and 
University of Ljubljana, Faculty of Mathematics and Physics, 1000, Ljubljana, Slovenia
\and 
Center of Excellence SPACE-SI, A\v{s}ker\v{c}eva cesta 12, 1000, Ljubljana, Slovenia
\and 
Department of Physics and Astronomy ``G. Galilei'', Padova University, Vicolo dell'Osservatorio 3, 35122 Padova, Italy
\and 
INAF - Padova Astronomical Observatory, Vicolo dell'Osservatorio 5, 35122 Padova, Italy 
\and 
Observatoire astronomique de Strasbourg 11 rue de l'Université, 67000 Strasbourg, France
\and 
Sydney Institute for Astronomy, University of Sydney, NSW 2006, Australia
\and 
Jeremiah Horrocks Institute, University of Central Lancashire, Preston, PR1 2HE, UK 
\and 
Monash Centre for Astrophysics, Monash University, Clayton, 3800, Australia
\and 
Institute of Astronomy, Cambridge University, Madingley Road, Cambridge CB3 0HA
\and 
University of Victoria, Department of Physics and Astronomy, Victoria, BC Canada V8P 5C2
\and 
Department of Physics and Astronomy, Macquarie University, NSW 2109, Australia
\and 
Macquarie research centre in Astronomy, Astrophysics and Astrophotonics, Macquarie University NSW 2109, Australia
\and 
Australian Astronomical Observatory, PO Box 296, Epping, NSW 2121, Australia
\and 
Leibniz-Institut fur Astrophysik Potsdam (AIP), An der Sternwarte 16, 14482 Potsdam, Germany
}
\date{Accepted for pubblication on A\&A}
\titlerunning{Thin disk with RAVE}
\authorrunning{S.\ Pasetto et al.}

\abstract
{}
{We study the Milky Way thin disk with the Radial Velocity Experiment (RAVE) survey. We consider the thin and thick
disks as different Galactic components and present a technique to statistically disentangle the two populations. Then we focus our attention on the thin disk component.}
{We disentangle the thin disk component from a mixture of the thin and thick disks using a data set providing radial velocities, proper motions, and photometrically determined distances.}
{We present the trend of the velocity dispersions in the thin disk component of the Milky Way (MW) in the radial direction above and below the Galactic plane using data from the RAdial Velocity Experiment (RAVE). The selected sample is a limited subsample from the entire RAVE catalogue, roughly mapping up to 500 pc above and below the Galactic plane, a few degrees in azimuthal direction and covering a radial extension of 2.0 kpc around the solar position. The solar motion relative to the local standard of rest is also re-determined with the isolated thin disk component. Major results are the trend of the velocity mean and dispersion in the radial and vertical direction. In addition the azimuthal components of the solar motion relative to the local standard of rest and the velocity dispersion are discussed.}
{} 

\maketitle

\section{Introduction}\label{Introduction}
The study of disk galaxies has played a fundamental role in the attempt to understand the intricate mechanisms that govern galaxy evolution \citep[see, e.g.,][for a recent review]{2011ARA&A..49..301V}.  The complex challenge of their origin and evolution has occupied the interests of the astronomical community for a long time \citep[e.g.,][]{2000dyga.book.....B, 1998gaas.book.....B, 1995gaco.book.....C, 1987gady.book.....B, 1985gpsg.book.....S}. However, the global picture of the origin and evolution of spiral galaxies requires thorough comprehension of the structure and evolution of our own Galaxy, where many of the relevant properties and processes can be studied in detail. In this process, by mapping the Galactic disk distribution,  obtaining a detailed description of our Galactic disk kinematics is of paramount importance in order to determine the structure and the evolution of the Milky Way.

The first fundamental steps in this process were done by the determination of the characteristic parameters of the disk through the study of the star-count equation \citep[e.g.,][]{1953stas.book.....T} as in the pioneering  models of \citet{1980ApJS...44...73B} or \citet{1983MNRAS.202.1025G}. These studies opened the way for more complete investigations by taking into account also the kinematic component of the data \citep[e.g.,][]{1987A&A...180...94B, 1996ASPC..102..345M}. As a consequence of the constantly growing amount of available data and our improving knowledge of Galactic structure, the initial ideas of galaxy formation \citep[e.g.,][]{1962ApJ...136..748E, 1978ApJ...225..357S, 1967ApJ...147..868P} have developed dramatically and merged with cosmological research becoming now either a part of a standard textbook picture \citep[e.g.,][]{1998gaas.book.....B, 2006ima..book.....C} or a current hot topic of research.

Continuing this long line of research the RAdial Velocity Experiment project (RAVE) was designed \citep[see][]{2008AJ....136..421Z, 2006AJ....132.1645S}: a stellar survey that provides spectroscopic parameters for stars of the Galaxy that are mainly part of the thin disk, thick disk or halo component. The power of this survey has been repeatedly demonstrated with a series of papers focused on the principal target of the survey, i.e. the study of the Galactic structure and kinematics  \citep[e.g., ][]{2008A&A...480..753V, 2011ApJ...728....7C, 2008MNRAS.391..793S, 2011MNRAS.418.2459H, 2011MNRAS.412.1237C, 2011MNRAS.412.2026S, 2011MNRAS.413.2235W, 2012NewA...17...22K}, identifying stellar streams \citep[e.g.,][]{2008ApJ...685..261K, 2008MNRAS.384...11S, 2011ApJ...728..102W, 2011MNRAS.411..117K} and studying the chemistry of Milky Way (MW) components \citep[e.g.,][]{2010ApJ...721L..92R, 2011ApJ...743..107R, 2010ApJ...724L.104F, 2011AJ....142..193B, 2012MNRAS.419.2844C}.

In this paper we proceed with the general efforts devoted to understanding the properties of the Galactic disk from a kinematic point of view. Ultimately, a large spectroscopic sample of disk stars can reveal the characteristic parameters of the structure of our Galactic disk, the evolutionary history of the disk populations and the physical processes that acted on the stars in order to produce the picture that the RAVE data show us today.

Our task is to attempt to break the parameter degeneracy that results from the imprecise (or incomplete) knowledge of the observed parameters for the catalogue stars. The RAVE survey produces directly several phase space parameters. We will directly work with: the Galactic coordinates $\left( {l,b} \right)$, the radial velocities $v_r  \equiv \left\langle {v_{{\rm{hel}}} ,{\bf{\hat r}}} \right\rangle {\bf{\hat r}}$, where the unitary vector to a generic star is a function of the directions ${\bf{\hat r}} = {\bf{\hat r}}\left( {l,b} \right)$.
Within their errors these parameters permit us to constrain the true velocity vector once the distance is known. All these parameters come, of course, with their uncertainties. In particular, with indirect methods the distances ${\bf{r}}_{{\rm{hel}}}  = r_{{\rm{hel}}} {\bf{\hat r}}$
 of the stars can be derived only within a given error \citep[e.g.,][]{2010A&A...522A..54Z}.

Although these quantities, if complemented by accurate distances and proper motions would provide the full phase space parameters for each star, i.e.,  the initial conditions for a complete N-body integration or equivalently the boundary conditions for the Liouville equation, there are several reasons that prevent us from following this direct approach: 
\begin{itemize}
	\item the error bars of the above mentioned determinations leave too large a space of possible solutions, which prevents us from achieving an acceptable convergence of the results;
	\item the proximity of the stars in our sample prevents us from really following the trajectory of each star, not permitting us to evaluate the exact behaviour of more distant parts of the Galaxy.
\end{itemize}

Thus, we are in the apparently paradox situation that successful methods for external galaxies do not yield the same quality of results in our much closer Galaxy (see e.g., the results based on the full N-body integrations \citep{1997A&A...327..983F}, on Schwarschild methods or their more recent evolution, the Made to Measure methods e.g., by \citet{2009MNRAS.395.1079D}). The Galaxy description we are going to utilize here is only of statistical nature, and it is based on the collisionless Boltzmann approximation to describe the Galaxy as a set of two components (we are considering only thin and thick disk stars in our description). This will prevent us from easily deriving any consideration of dynamical nature, and we will limit ourselves to a simpler kinematic description. In particular, here we present a statistically determined description of the thin disk component alone that can be achieved, within reasonable errors, in the solar neighbourhood.

Nonetheless, the problem of extending the kinematic parameters estimated here for the solar neighbourhood to the entire disk can be justified based on symmetry considerations and on the comparison with external spiral galaxies. If we limit ourselves to a kinematic description, i.e., a steady state description for a one-component disk (namely the thin disk), then the Jeans theorem \citep[e.g.,][]{1987gady.book.....B} authorizes us to look at the solar vicinity as a paradigm of the Galactic disk kinematics \citep[e.g.,][]{1986MNRAS.221..857M} as long as we do not attempt too detailed a description, \citep[e.g.,][]{2011MNRAS.417..762Q}. We will extensively discuss the limitation of our approach in later sections of our paper.

The structure of this paper is the following: in Sect. \ref{Methodologyadopted}, we review the hypotheses of the methodology adopted and the data selection criteria. In the Sect. \ref{results} we present the results and in Section 4 our conclusions.

\section{Methodology adopted}\label{Methodologyadopted}
\subsection{Target}\label{Targhet}  
In this paper, our target is to present the trend of the first two moments of the velocity distribution (the mean and the central moment of order 2) for the thin disk component along the meridional plane $\left( {O,R,z} \right)$
 of the MW. Using a methodology recently developed in \citet[][]{GreenPaper} (hereafter Paper I) we are able to disentangle the thin disk and thick disk components from the mixture of the two disk populations near the sun. Then, with the help of proper motions and distances photometrically determined by Zwitter et al. (2010) we can follow the trend of the thin disk velocity ellipsoid described in cylindrical coordinates $\left( {O,R,\varphi ,z} \right)$.
 
\subsection{Data selection}\label{Dataselection}
We review here the data selection, while the methodology is fully explained in Paper I. We used an internal release of the RAVE data catalogue equipped with photometrically determined distances (Zwitter et al. 2010), radial velocities and proper motions. This provides us with a full phase space description for about 210,000 stars consisting of a mixture of thin and thick disk and halo stars.

\begin{figure}
\resizebox{\hsize}{!}{\includegraphics{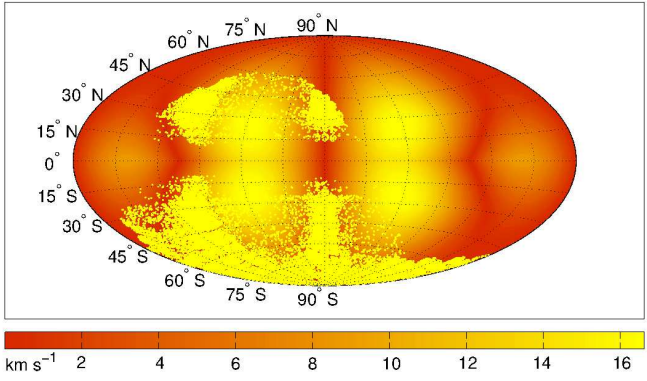}}
\caption{
Celestial sphere distribution in an Aitoff projection of the selected stellar sample (yellow dots) as outlined in the text. The background density map is explained in the text.}
\label{SkyDist}
\end{figure}

In order to limit our sample only to the thin \textit{and }thick disk components, we select from the RAVE catalogue all the stars with $\log _{10} g > 3.5$ (with $g$ denoting the surface gravity of the star), the near-infrared colours from the Two Micron All Sky Survey (2MASS), \citep{2006AJ....131.1163S} $J - K \in \left[ {0.3,1.1} \right]$ mag in the $J$ and $K$ band (see \cite{2006AJ....132.1645S} and \cite{2008AJ....136..421Z} for details), the latitude of the sample $\left| b \right| > 10$
deg, $\left[ {\text{Fe}/\text{H}} \right] >  - 1$ (from \cite{2008AJ....136..421Z} in order to minimize any halo contamination), $\left\| {{\mathbf{v}}_{{\rm{hel}}} } \right\| < 150{\rm{km s}}^{ - 1} $ as absolute value of the true velocity vector and $S/N > 20$ for the signal-to-noise ratio of the spectra. Finally, we cut the sample with the kinematical condition investigated in Paper I: $\Delta v_{r}^{\text{RAVE}}\left( {{l}_{i}},{{b}_{i}} \right)>2.0 \delta v_{r}^{\text{G}}\left( {{l}_{i}},{{b}_{i}} \right)$  (Eqn. (6) in Paper I) where $\delta v_{r}^{\text{G}}\left( {{l}_{i}},{{b}_{i}} \right)$ is the difference between a maximum value $\max \left( {v_r^{\rm{G}} } \right)$ and a minimum value $\min \left( {v_r^{\rm{G}} } \right)$ of the Galactic component of the radial velocity in the direction of the star $s_i$ where $r \in \left[ {0,1} \right]$ kpc. This condition has been investigated in Paper I in order to reduce the indetermination of the kinematical parameters for the most distant stars and the large scale effects along the selection process. An isocontour plot of this inequality based on the assumed Galactic gravitational potential field \citep[e.g.,][]{2003ASPC..298..153B,2006A&A...451..125V} is given in the background contour plot in Fig. \ref{SkyDist} and can be directly compared with Fig. 1 of Paper I. As we can see from this contour
plot, the highest $\delta v_{r}^{\text{G}}\left( {{l}_{i}},{{b}_{i}} \right)$ (yellowish zones) occur at low latitude and the smallest in the red zones. For $b=0$ $\delta v_{r}^{\text{G}}\left( {{l}_{i}},{{b}_{i}} \right)$ is the terminal velocity relation and we see that the the largest induced error is of the order of 16 km s$^{-1}$ at a distance of 1 kpc and at low Galactic latitudes not covered by the RAVE survey.  Here we assume that the true velocity vector is defined in a suitable reference system for the velocity space $\left( {O,U,V,W} \right)$  aligned with the configuration reference system at the solar position. The $U$ direction points to the Galactic coordinates $\left( {l,b} \right) = \left( {180,0} \right)$ deg, $V$ runs through $\left( {l,b} \right) = \left( {90^\circ ,0} \right)$ deg and $W$ points to the North Galactic Pole (NGP).

This data selection leaves us with a sample of 38,805 dwarf stars that map the solar neighbourhood's stellar distribution avoiding the influence from the small halo component (see Figure \ref{SkyDist}). 

\subsection{Disentangling technique}\label{Disentanglingtechnique}
In the classical kinematic description, a single collisionless stellar population satisfies the collisionless Boltzmann equation, i.e., a linear and homogeneous partial differential equation in the phase space distribution function (DF), $f$ (once the total potential $\Phi _{\text{tot}} $ of the whole Galaxy is assumed). Since this equation is linear in $f$, the whole system can be described as a mixture of populations, e.g., $f = \sum\limits_i {f_i } $.  In Paper I (to which we refer the reader for further details) a methodology based on the Singular Value Decomposition solution for overdetermined systems was outlined in order to statistically derive the first moments of a DF $f_{{\rm{mix}}}  = f_{{\rm{thin}}}  + f_{{\rm{thick}}} $ for a stellar mixture of thin and thick disks. Then, we statistically disentangled the thick disk population alone, assuming a generalized ellipsoidal DF, $f_{\text{thick}} $, in modification of the original work of \citet{2004A&A...427..131C}.

The selection procedure works only on the radial velocity. At every iteration, say $j$, by randomly choosing a radial velocity value within its error bar, say $v_r^{[j]}(s_i)$ for each star $s_i  \in P^{[j]}$, i.e. a star that passed our selection criteria in Section \ref{Dataselection}, we are left with a selected RAVE sample of stars $P^{[j]}$ within their associated errors:
\begin{equation}
	P^{[j]}  \equiv \left\{ {s_i |v_r^{[j]} \left( {s_i } \right) \in \left[ {v_r \left( {s_i } \right) - \Delta v_r \left( {s_i } \right),v_r \left( {s_i } \right) + \Delta v_r \left( {s_i } \right)} \right]} \right\}
\end{equation}
The procedure then isolates the thick disk component by randomly excluding stars in order to maximize a statistical weight parameter $q$ described in Paper I \citep[see also][]{2004A&A...427..131C}. Once the population that maximizes $q$
 within the machine error precision (namely $q = \infty $) has been isolated, we are naturally left with two samples. The sample $S^{[j]} $, with  $S^{[j]}  \subset P^{[j]} \forall j$, describing the thick disk and the sample $P^{[j]} \backslash S^{[j]} $ or $\overline {S^{[j]} } $ (read "not $P^{[j]} $") describing the thin disk, is defined as: 
\begin{equation}	
	T^{[j]}  \equiv P^{[j]} \backslash S^{[j]}  = \left\{ {s_i |s_i  \in P^{[j]}  \wedge s_i  \notin S^{[j]} } \right\}\forall j
\end{equation}
thus once the thick disk sequence of stars for the iteration $j$ is removed all the other stars are considered to be thin disk stars. 
We explicitly state that:
\begin{itemize}
	\item we do not limit the determination of the errors to a preselected sample of stars. For instance, a  generic star $s_i$ of $P^{[j]}$ can belong to the sample $S^{[j]}$ for the iteration number $j$ and to the sample $T^{[j]}$ for the iteration number $j+1$. The indices $j$ and $i$ are not correlated.
	\item the \textit{thin }disk selection is based only on the radial velocities and their errors once the sample $P$ has been chosen. This makes use of the good quality errors as determined by the RAVE survey in Eqn. 1.
\end{itemize}

We emphasize that the first point is fundamental: we are not assigning any probability weight to a single star in the sense of being a thin disk or thick disk star, \emph{even though} we know its full phase space description. This assigns much more generality to our method and prevents us from biasing the results.

\begin{figure}
\resizebox{\hsize}{!}{\includegraphics{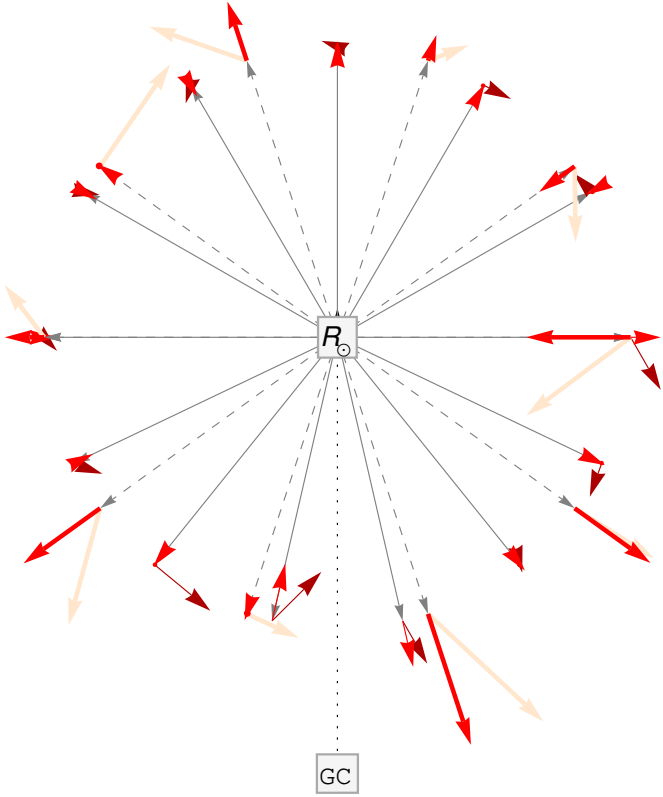}}
\caption{
Sketch of the geometry exploited by the method. The observer is located at the Sun's position $R_{\odot}$ at the centre. The Galactic centre (GC) is at the bottom of the figure. The directions on the celestial sphere to each star $s_i$ are indicated with dashed arrows for thick disk stars (the sample $S^{[j]}$) and solid arrows for the thin disk stars (the sample $T^{[j]}$). The red thick arrows represent the radial velocities of the thick disk, whose corresponding unknown true velocity vector is in light red. The thin disk radial velocities are represented by the thin red arrows, and the corresponding unknown dark red arrows represent the true velocity vectors. Note that the dark red arrows are generally longer in the GC direction and smaller in the anti-Galactic centre direction as expected from the velocity dispersion trend. }
\label{Frecce}
\end{figure}

A schematic view of how the method works is sketched in Figure \ref{Frecce} where for a generic sample $P^{[j]} $, the direction to each star, ${\bf{\hat r}}$, is represented by grey arrows (solid for the thin disk stars and dashed for the thick disk). The true, \textit{unknown} velocity vectors for the thick disk stars ${\bf{v}}_{{\rm{hel}}}^{{\rm{thick}}} $, drawn from the DF $f_{{\rm{thick}}} $, are represented in light red. The sketch shows the thin disk star velocity vectors, ${\bf{v}}_{{\rm{hel}}}^{{\rm{thin}}} $, with dark red arrows that are slightly greater closer to the Galactic centre (GC) and smaller in the radial outward direction for $R > R_ \odot  $, with $R_ \odot  $ being the solar position. This is an expected trend for the thin disk velocity ellipsoid which is supposed to have larger ellipsoidal axes in the inner zones of the Galaxy and smaller ones outside \citep[e.g.,][]{1998gaas.book.....B}. The radial projection of each velocity vector is shown with red arrows: thinner for the thin disk and thicker for the thick disk. We used in the selection process only the radial velocity (red arrows) ${\bf{v}}_\parallel   = \left\langle {{\bf{v}}_{{\rm{hel}}} ,{\bf{\hat r}}} \right\rangle {\bf{\hat r}}$
 provided by RAVE, and thanks to the different velocity dispersion of the thin and thick disk components (graphically realized with longer and shorter red arrows) we managed to disentangle the thick disk component as in Paper I. Once the thick disk stars have been selected (dashed grey arrows) we call all the other stars ``thin disk stars'' (the solid grey arrows). We determine their true velocity vectors (dark red arrows) by making use of the proper motion values 
tabulated in the RAVE internal data release and
calibrated with photometrically determined distances from Zwitter et al. (2010). In addition we use the matrix transformation of \citet{1987AJ.....93..864J} corrected to J2000 as in \citet{2003A&A...405..931P} in order to derive the velocity dispersion tensor for the thin disk alone.

\section{Results}\label{results}
We present now the analysis of the averaged results for a large set of statistical realizations of the sample $T^{[j]}$ for the single thin disk component (typically $j=1,..,N$ with $N>1024$) as outlined in the previous section. With the concept of a \textit{single component} we mean each part of the system that is characterized by a specific property, i.e. the chemical elements $\mathbb{C} $, the age $t$, the mass-to-light ratio ${{M \mathord{\left/
 {\vphantom {M L}} \right.
 \kern-\nulldelimiterspace} L}}$ in some specific band, the kinematic parameters, the scale length, etc. If we collectively call $\Xi$ the sample of the physical properties that we are \textit{not} interested in, its cardinality (i.e. the number of its elements) can be very high, or equal to the cardinality of the continuum. We will average the DF over all the properties we do not wish to study so that we will refer to the thin (or thick disk) DF simply as $f_{{\text{thin}}}  \propto \int_{}^{} {f_{{\text{thin}}} d\Xi } $. Since the mapping of the meridional plane kinematics is our principal goal, we will limit ourselves to the kinematic description of the thin disk DF (through only its first and second velocity moment). Thus once the sample of stars $T^{[j]}$ is selected, we are left with the proper motion and their errors, radial velocities and their errors, distances and their errors $\forall s_i  \in T^{[j]}$ (and the sky directions considered without errors).

\subsection{Data binning }
Once the thin disk velocity vectors ${\bf{v}}_{{\rm{hel}}} $ have been found, we proceed by binning the sample in the meridional plane. We are not binning in the $\phi $ direction in order to exploit the axisymmetric assumption we made when the selection procedure was developed as explained in Paper I, i.e. we are summing up all the $\phi _{s_i } $ for each star $s_i$ and the Sun at the azimuthal position $\phi _ \odot   = 0$ in a Galactocentric cylindrical reference system. The choice of the bin size does not affect the results we are presenting once a sufficiently large bin (say, $\geqslant 100$ pc) is adopted since we are interested in the general trend of a smooth function of the independent variables $R$ and $z$ (and not in the description of the granularity of the DF). Different statistical approaches can be shown to give the same physical results, e.g., the statistical estimators in \cite{1994A&A...289..640V}.

\subsection{The mean motion}

\begin{table*}
\caption{Trend of the mean radial velocity $\bar v_R $ expressed in $[{\rm{\text{km s}}}^{ - 1}] $ and the distance from the Galactic plane bins are in [kpc]. Empty slots indicate parameter ranges with too few stars.}
 \centerline {\begin{tabular}{|c|r|r|r|r|r|r|}
 \hline
$\bar v_R (R,z)$&\textbf{$<$-0.6}&\textbf{]-0.5,-0.3]}&\textbf{]-0.3,-0.1]}&\textbf{]-0.1,0.1]}&\textbf{]0.1,0.3]}&\textbf{]0.3,0.5]}  \\
 \hline 
\textbf{7.6}&  & $21.7\pm50.0$ &   &   & $63.1\pm0.0$ &  \\
\textbf{7.8}&$35.3\pm67.2$ & $14.2\pm12.1$ & $12.8\pm28.0$ &   & $8.2\pm21.3$ & $17.5\pm20.1$\\
\textbf{8.0}&$3.7\pm6.6$ & $16.4\pm7.5$ & $9.1\pm4.3$ &   & $13.2\pm9.3$ & $14.4\pm9.6$\\
\textbf{8.2}&$20.4\pm24.5$ & $14.9\pm4.4$ & $9.9\pm2.6$ & $10.0\pm11.9$ & $13.3\pm4.5$ & $16.4\pm8.2$\\
\textbf{8.4}&$19.2\pm18.7$ & $11.4\pm3.1$ & $10.9\pm2.7$ & $10.9\pm1.0$ & $13.4\pm2.6$ & $12.5\pm4.9$\\
\textbf{8.6}&  & $10.4\pm3.1$ & $10.2\pm1.6$ & $10.1\pm2.6$ & $11.5\pm2.4$ & $5.7\pm2.6$\\
\textbf{8.8}&$6.5\pm11.6$ & $6.1\pm3.1$ & $7.4\pm2.2$ & $17.2\pm45.3$ & $6.4\pm4.0$ & $8.0\pm6.5$\\
\textbf{9.0}&$-6.2\pm31.7$ & $6.7\pm3.8$ & $10.3\pm10.9$ &   & $9.4\pm26.0$ & $4.1\pm3.5$\\
\textbf{9.2}&  & $-3.8\pm10.6$ & $12.9\pm0.0$ &   &   &  \\ 
\textbf{9.2}&  & $22.4\pm0.0$ &   &   &   &  \\ 
\hline
\end{tabular}}
\label{Tabella01vm}
\end{table*}

\begin{table*}
\caption{For completemess we present here the trend of the mean circular velocity $\bar v_{\phi} $ expressed in $[{\rm{\text{km s}}}^{ - 1}] $ and the distance from the Galactic plane bins are in [kpc].  Empty slots indicate parameter ranges with too few stars. Refer to the text for the limitations of our approach along this direction.}
 \centerline {\begin{tabular}{|c|r|r|r|r|r|r|}
 \hline
$\bar v_{\phi} (R,z)$&\textbf{$<$-0.6}&\textbf{]-0.5,-0.3]}&\textbf{]-0.3,-0.1]}&\textbf{]-0.1,0.1]}&\textbf{]0.1,0.3]}&\textbf{]0.3,0.5]}  \\
 \hline 
\textbf{7.6}& & $6.6\pm17.1$ &   &   &  & \\
\textbf{7.8}&$-43.1\pm78.5$ & $-12.5\pm8.3$ & $12.5\pm26.2$ &   & $2.7\pm7.7$ & $2.2\pm3.6 $\\
\textbf{8.0}&$-47.2\pm61.9$ & $-10.8\pm3.9$ & $2.5\pm1.6$ &   & $5.8\pm4.6$ & $-8.0\pm3.0$\\
\textbf{8.2}&$-42.6\pm45.9$ & $-4.5\pm1.2$ & $6.0\pm1.6$ & $7.8\pm8.8$ & $6.1\pm2.9$ & $-8.6\pm2.8$\\
\textbf{8.4}&$-36.6\pm41.0$ &  & $8.3\pm1.5$ & $11.3\pm1.5$ & $7.2\pm0.9$ & $-5.7\pm2.0$\\
\textbf{8.6}&$-42.6\pm58.0$ & $-2.7\pm0.5$ & $9.9\pm2.6$ & $13.1\pm2.9$ & $9.2\pm2.2$ & $-7.8\pm2.9$\\
\textbf{8.8}&$-44.7\pm140.4$ & $-3.0\pm0.6$ & $9.5\pm4.3$ & $9.2\pm29.6$ & $7.2\pm5.8$ & $-11.2\pm8.4$\\
\textbf{9.0}&$-29.6\pm103.9$ & $-5.8\pm4.1$ & $5.0\pm7.7$ &   & $5.5\pm20.9$ & $-11.7\pm24.9$\\
\textbf{9.2}& & $-2.6\pm4.3$ &  &   &   & \\
\hline
\end{tabular}}
\label{Tabella02vm}
\end{table*}

\begin{table*}
\caption{Trend of the mean vertical velocity $\bar v_z $ expressed in $[{\rm{\text{km s}}}^{ - 1}] $ and the distance bins from the Galactic plane are in [kpc].  Empty slots indicate parameter ranges with too few stars.}
 \centerline {\begin{tabular}{|c|r|r|r|r|r|r|}
 \hline
$\bar v_z (R.z)$&\textbf{$<$-0.6}&\textbf{]-0.5.-0.3]}&\textbf{]-0.3.-0.1]}&\textbf{]-0.1.0.1]}&\textbf{]0.1.0.3]}&\textbf{]0.3.0.5]}  \\
 \hline 
\textbf{7.6}& & $18.8\pm45.2$ &   &   &  &  \\
\textbf{7.8}&  & $15.6\pm13.1$ & $15.1\pm32.7$ &   & $22.0\pm56.6$ & $10.8\pm13.3$ \\
\textbf{8.0}&$14.9\pm20.3$ & $13.5\pm7.3$ & $10.3\pm6.5$ &   & $6.8\pm4.5$ & $10.5\pm5.8$\\
\textbf{8.2}&$18.5\pm22.1$ & $10.3\pm4.2$ & $9.0\pm2.0$ & $5.4\pm7.9$ & $8.8\pm2.3$ & $10.0\pm4.1$\\
\textbf{8.4}&$13.4\pm15.5$ & $10.7\pm4.3$ & $8.1\pm1.2$ & $7.2\pm1.3$ & $9.2\pm2.7$ & $13.4\pm7.2$\\
\textbf{8.6}&$16.2\pm23.6$ & $9.9\pm3.7$ & $9.4\pm2.0$ & $8.1\pm3.1$ & $10.3\pm3.7$ & $16.2\pm8.8$\\
\textbf{8.8}&$5.1\pm16.5$ & $8.2\pm3.6$ & $10.8\pm4.3$ & $10.0\pm33.3$ & $12.5\pm11.8$ & $17.8\pm18.0$\\
\textbf{9.0}&$21.3\pm75.9$ & $9.6\pm9.4$ & $12.3\pm19.5$ &   & $5.2\pm20.7$ & $11.2\pm26.2$\\
\textbf{9.2}& & $25.8\pm47.1$ & &   &   & \\
\hline
\end{tabular}}
\label{Tabella03vm}
\end{table*}

In the Tables \ref{Tabella01vm}, \ref{Tabella02vm}, and \ref{Tabella03vm} the velocity components of the mean velocity vector are listed along with their errors for a selected sample of bins. A common trend is present in the values of the tables, although it is also evident that the large error bars prevent us from obtaining a stable fitting function. This means that although the average values for $\bar v_R $, $\bar v_\phi  $
 and $\bar v_z $ show a smooth trend, we will not be in the position to compute their derivatives $\partial _R \bar v_i $
 or $\partial _z \bar v_i $ which remain a primary target of forthcoming surveys such as Gaia. In general the tables show a lack of data in the Galactic centre and anticentre directions as a consequence of the selection procedure outlined in Section \ref{Dataselection} and the Galactic latitude selection cuts ($\left| b \right| > 10$ deg). Nevertheless, when possible, we retain values as e.g., $\bar v_R= 18.8 \pm 45.2$ ${\rm{km}}\;{\rm{s}}^{ - 1} $ in Table \ref{Tabella01vm} for the distance range $z \in \left] { - 0.5, - 0.3} \right]$
kpc (and similar in Tables \ref{Tabella02vm} and \ref{Tabella03vm}), because their mean values are in agreement with the general trend of the analysed property, e.g., $\bar v_R$, expressed by the neighbouring bins in the Tables.

\subsubsection{Mean radial motion}\label{Meanradialmotion}
The trend of the mean radial motion $\bar v_R $ is presented in Table \ref{Tabella01vm}. The column describing the vertical range $z \in \left] { - 0.1,0.1} \right]$ kpc contains the closest and our best-determined values in the solar neighbourhood.  Assuming the location of the Sun to be in the range $R \in \left] {8.4,8.6} \right]$ kpc we recover a value of $\bar v_R = 10.9 \pm 1.0\;{\rm{km}}\;{\rm{s}}^{ - 1} $, in general agreement with the recent independent determination \citep[e.g.,][]{2009MNRAS.397.1286A} but with a larger error bar due to the statistical approach adopted. This value is in good agreement also with our recent determination deduced from the thick disk component alone in Paper I. Also, the value of $\bar v_R $ seems to be robust against possible large scale effect correction \citep[see also,][]{2012arXiv1207.3079S}. Table \ref{Tabella01vm} provide us with a consistent value for the LSR radial motion not only within the closest radial bin around the solar location, but also in its extended neighbourhood. Table \ref{Tabella01vm} shows no net global motion within the error bars, i.e., the values  show no net global radial motion of the stars with respect to the motion of the Sun. 
This is an indication of the absence of a net radial migration trend with respect to the solar motion and is in line with results already obtained much closer inside and further outside of our range of distances in the milestone paper of \citet{1989AJ.....97..139L}. This means that for a mixed set of stellar populations, the assumption of a zero net expansion is quite reasonable within the error bars even if a small outward trend can be evinced from the plot. 
On the plane $z = 0$, the radial motion of the stars at the solar position remains essentially unchanged from $ - 0.5$
kpc to $ + 0.5$ kpc. 
Similar considerations hold along all radial extensions of the RAVE data sample within or outside the solar radial position 
for above and below the plane respectively.

\subsubsection{Mean circular  motion}\label{Meancircularmotion}
The values for the mean circular motion $\bar v_\phi $, are presented in Table \ref{Tabella02vm}. The solar neighbourhood value for this component is $v_{V, \odot }  = 11.3 \pm 1.5\;{\rm{km}}\;{\rm{s}}^{ - 1} $, which is slightly higher than the classical literature values \citep[e.g.,][]{1998gaas.book.....B} but agrees with recent values such as in \citet{2006A&A...445..545P} and \citet{2010MNRAS.403.1829S}. Nevertheless, this value is slightly more difficult to estimate because of the superposition of several effects: the Galactic rotation, the peculiar velocity of the Sun relative to the local standard of rest at the solar position (LSR), the distribution of velocities of the sample of stars, and finally the presence of the thick disk stars. Here we have statistically excluded the thick disk's kinematic contribution to this value. The relation between the other contributors, for a single component thin disk model, can be approximated with the help of the Jeans equation as
\begin{equation}\label{EQnVC}
	\begin{array}{c}
	 \left| {{\bf{\bar v}}_c \left( {R,z} \right)} \right| = \left[ {V_{{\rm{LSR}}}^2  - \frac{{\partial \ln \rho }}{{\partial \ln R}}\left( {\sigma _{{\rm{RR}}}^2  + \sigma _{{\rm{Rz}}}^2 } \right) + \left( {\sigma _{{\rm{RR}}}^2  + \sigma _{\phi \phi }^2 } \right)} \right. \\ 
	 \left. { + R\left( {\frac{{\partial \sigma _{RR}^2 }}{{\partial R}} + \frac{{\partial \sigma _{Rz}^2 }}{{\partial z}} + \frac{{\partial \Phi _{\text{tot}} }}{{\partial R}}} \right)} \right] \\ 
	 \end{array}
\end{equation}
with $V_{LSR}^2 \left( R \right)$ as the velocity of the circular orbit at the radius $R$, $\rho$ as the stellar population density profile and $\sigma_{ij}$ the velocity dispersion tensor elements introduced below. This equation shows how part of the kinetic energy of any disk population goes into rotation and part in the velocity dispersion, which is necessary to stabilize the disk by achieving a suitable $Q$ parameter \citep[after][]{1964ApJ...139.1217T}. 
Nevertheless, there are a few important reasons why we cannot easily exploit the previous equation in order to disentangle the different contributions to the mean circular velocity:
\begin{enumerate}
	\item This result is an expected shortcoming of the methodology adopted to disentangle the thin and thick disk components. As already explained in Paper I, although the collisionless Boltzmann equation is linear in the phase space distribution function, the velocity dispersion moments become linear \textit{if and only if }the mean velocities of the partial distribution have the same mean (i.e., in our case, we can correct the thin disk and thick disk only in the radial and vertical directions). In the azimuthal direction clearly this is not the case (as evident from Table \ref{Tabella02vm}), preventing us from using the result on the dispersion tensor for the thick disk of Paper I in order to correctly recover the value $v_{V, \odot } $ from Equ. \eqref{EQnVC} and Tables \ref{tabsrr}, \ref{tabsrz} and \ref{tabsff}.
	\item  While the total potential $\Phi _{{\rm{\text{tot}}}} $ and the scale length, $h_R $, which enter in the density $\rho \left( {R,z} \right)$  distribution of the population, can be related thanks to the Poisson equation \citep[see, e.g.,][]{PhDThesis} or the Appendix of \citet{2006A&A...451..125V}, unfortunately the velocity dispersion errors presented below do not permit us to compute the terms $ \frac{{\partial \sigma _{RR}^2 }}{{\partial R}} $ and $\frac{{\partial \sigma _{Rz}^2 }}{{\partial z}}$. 
 \item The age-velocity dispersion relation implies that the spatial distribution and kinematics of the stars are strongly dependent on the age of the stars, i.e., the dispersion tensor has a strong dependence on the time $\sigma _{ij}  = \sigma _{ij} \left( {R,z,t} \right)$
as well as on the scale height $h_z  = h_z \left( t \right)$ \citep[e.g.,][]{1974HiA.....3..395W, 1977A&A....60..263W}. Hence, in equation \eqref{EQnVC} we can express the time dependence of ${\bf{\bar v}}_\phi  $ by writing ${\bf{\bar v}}_\phi  \left( {R,z,t} \right) = {\bf{\bar v}}_\phi  \left( {R,z,\sigma _{ij} \left( {R,z,t} \right)} \right)$ or more briefly ${\bf{\bar v}}_\phi   = {\bf{\bar v}}_\phi  \left( {\sigma _{ij} \left( {R,z,t} \right)} \right)$ to show that if we assume $\frac{{\partial \Phi _{{\rm{\text{tot}}}} }}{{\partial t}} \ll \frac{{\partial \sigma _{ij} }}{{\partial t}}$ then the dominant time dependence in ${\bf{\bar v}}_\phi  $
 is given by $\sigma _{ij}  = \sigma _{ij} \left( {R,z,t} \right)$. In this case, it is convenient to extrapolate the value of the motion of the solar neighbourhood ${\bf{\bar v}}_\phi  $  as  ${\bf{\bar v}}_\phi   = \mathop {\lim }\limits_{t \to t_{\min } } {\bf{\bar v}}_\phi  \left( {\sigma _{ij} \left( {\bar R_ \odot  ,\bar z_ \odot  ,t} \right)} \right)$
 where $\left\{ {\bar R_ \odot  ,\bar z_ \odot  } \right\}$ then is the range $\left\{ {R,z} \right\} \in \left] {8.2,8.4} \right] \times \left] { - 0.1,0.1} \right]$
kpc of Table 2 and $t_{\min } $ is the minimal time to consider a young population in statistical collisionless equilibrium \citep[e.g.,][]{2009MNRAS.397.1286A}. Nevertheless, by applying such a time binning (on the age of the stars) we should limit ourselves to the immediate solar neighbourhood, namely only the closest bin $\left\{ {\bar R_ \odot  ,\bar z_ \odot  } \right\}$
 to deal with a suitable amount of stars per bin, which is not the goal of the present paper. Furthermore, if the origin of the age-velocity dispersion relation stands in the gravitational potential fluctuation caused by transient spiral structure \citep[e.g.,][]{2004MNRAS.350..627D} the approximation $\frac{{\partial \Phi _{{\rm{\text{tot}}}} }}{{\partial t}} \ll \frac{{\partial \sigma _{ij} }}{{\partial t}}$
 may fail and the previous argument for the use of Eqn. \eqref{EQnVC} does not hold any more.
\end{enumerate}

Nevertheless, we present these results in Table \ref{Tabella02vm} to be complete. We postpone the study of the trend of $\bar v_\phi$ in the meridional plane to future works and different techniques \citep[e.g.,][see also the conclusion in Section 4]{2012ApJ...747..101M}.

\subsubsection{Mean vertical motion}\label{Meanverticalmotion}
Finally, the mean vertical motion is presented in Table \ref{Tabella03vm}. The value around the solar position is $7.2 \pm 1.3\;{\rm{km}}\;{\rm{s}}^{ - 1} $
 in excellent agreement with the standard adopted literature values \citep[e.g.,][]{1998gaas.book.....B}. No apparent trend is evident within the errors, taking $\forall R$ and $\forall z$ into consideration.
 
\subsection{Velocity dispersion tensor}\label{Dispersionvelocitytensor} 
The trend of the velocity ellipsoid along the radial and vertical direction of the Galaxy is scarcely known due to the solar location within the Galaxy disk $z_ \odot   \cong 0$ and the resulting dust absorption  that affects the observations towards the Galactic centre. The pioneering work of \citet{1989AJ.....97..139L} remains as solitary attempt to observationally address this difficult task and we also will limit ourselves only to the close solar neighbourhood by neglecting the use of giant stars in this work.

Nevertheless, the RAVE survey has the advantage of letting us work with an extended number of stars, and to permit us to sample the trend of the thin disk velocity ellipsoid in a radial range of $R \in \left[ {7.6,9.2} \right]$ kpc from the Galactic centre, an extension that is scarcely sampled in other works such as \citet{1989AJ.....97..139L}.

We consider now the analysis of the terms $\sigma _{RR} $ and $\sigma _{zz} $. 

\begin{figure}
\resizebox{\hsize}{!}{\includegraphics{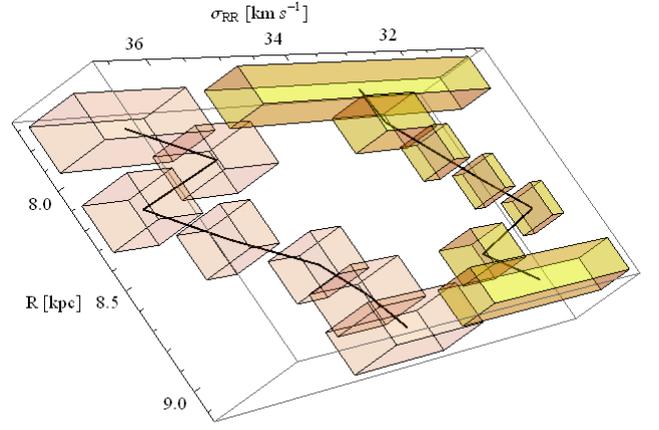}}
\caption{
An arbitrarily selected element and range from the $\sigma _{ij} $ tensor, here  $\sigma _{RR} $ as an example of the three-dimensional error box that surrounds every point for two different vertical ranges (corresponding to the two upper panels of Fig. \ref{Allplot} for the same velocity dispersion tensor). Black lines connect the centres of the boxes.}
\label{casino}
\end{figure}

We start by looking at Fig. \ref{casino} where we show the trend of $\sigma _{RR} \left( {R,z} \right)$
 with the Galactocentric distance $R$. The plot shows the trends of the data, but due to the overlapping of the error bars and the expected symmetry of the data above and below the plane caused by the proximity of the Sun to the ideal plane $z=0$, the plots with the error bars are generally crowded. We present an example of these plots only for illustrative purposes and only for the element $\sigma _{RR} $
 in order to give to the reader a feeling of the behaviour of the error bars. For each data point, the errors are given by a triplet of numbers, $\left\{ {\Delta R,\Delta z,\Delta \sigma _{ij} \left( {R,z} \right)} \right\}$, making a visualization hard to realize. Each point lies within a three-dimensional error cube whose size is mainly a consequence of the error propagation formula (an element by element explicit derivation can be found in the Appendix of \citet{2009AJ....137.4149F}) and it depends on the number of stars within the selected spatial range (which, in turn, is variable for each realization $P^{[j]} $).

In order to avoid realizing such a complicated three-dimensional plot, in the figures we are presenting for a generic element, say $\sigma _{RR} \left( {R,z} \right)$ in Fig. \ref{SRR01}, \ref{SRR02} and \ref{SRR00}, we are omitting the errors in the $z$
 direction. 
 
\begin{figure*}
\resizebox{\hsize}{!}{\includegraphics{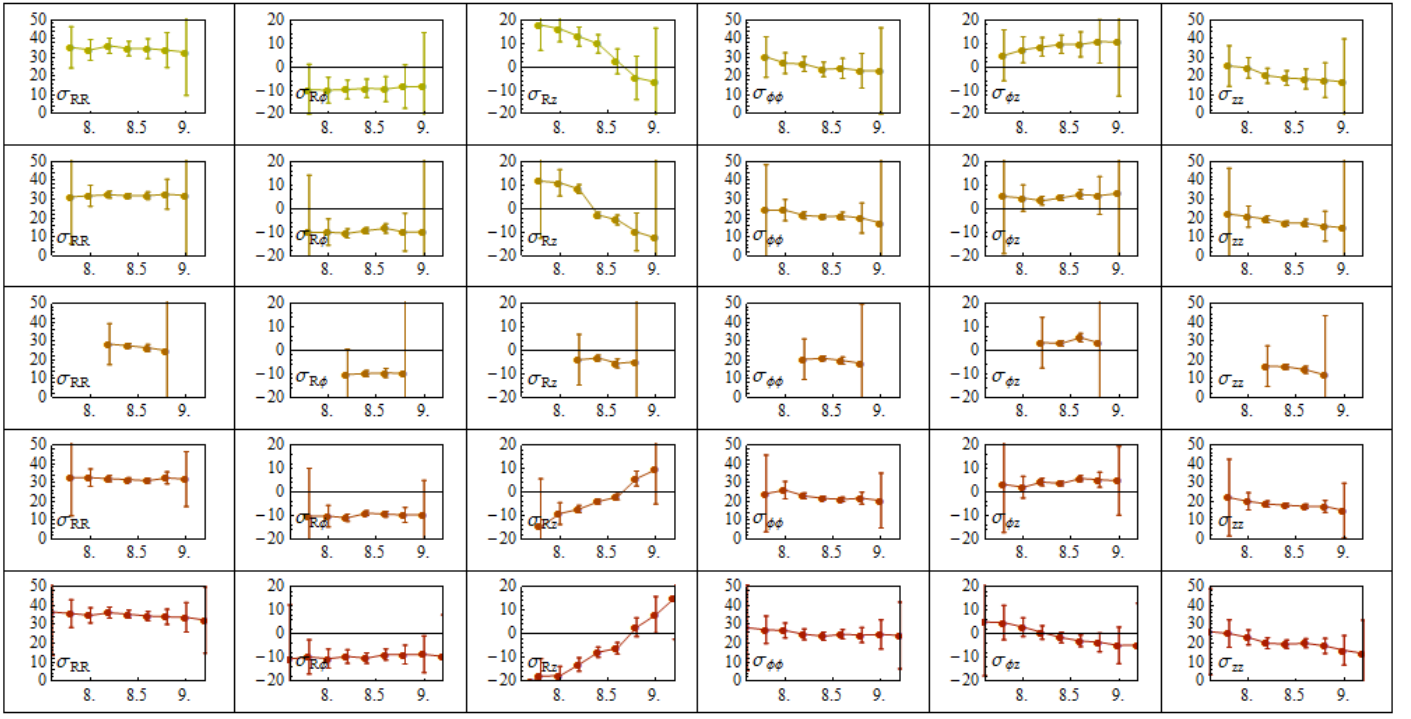}}
\caption{
Plot of all the velocity ellipsoid elements $\sigma_{ij}$ in the plane $(O,R,z)$. The axes are in [$\text{km s}^{-1}$] for the velocity dispersion (y-axes) and in [kpc] for the Galactocentric distances (x-axes). The axis labels are omitted in favour of a small indicator of the element we are considering in the lower left part of the plot.  In order to visualize different $z$ distance ranges, we use a gradation of increasing red for the plot lines from the upper row of plot, $z \in \left] {0.3,0.5} \right]{\rm{kpc}}$ down to the bottom row for the range $z \in \left] {-0.5,0.3} \right]{\rm{kpc}}$: the first row of the upper 6 panels refers to $z \in \left] {0.3,0.5} \right]{\rm{kpc}}$, the second to
 $z \in \left] {0.1,0.3} \right]{\rm{kpc}}$, the central to the Galactic plane $z \in \left] {-0.1,0.1} \right]{\rm{kpc}}$. The fourth row includes panels for $z \in \left] {-0.1,-0.3} \right]{\rm{kpc}}$ and the fifth one to $z \in \left] {-0.3,-0.5} \right]{\rm{kpc}}$.}
\label{Allplot}
\end{figure*}

\begin{figure}
\resizebox{\hsize}{!}{\includegraphics{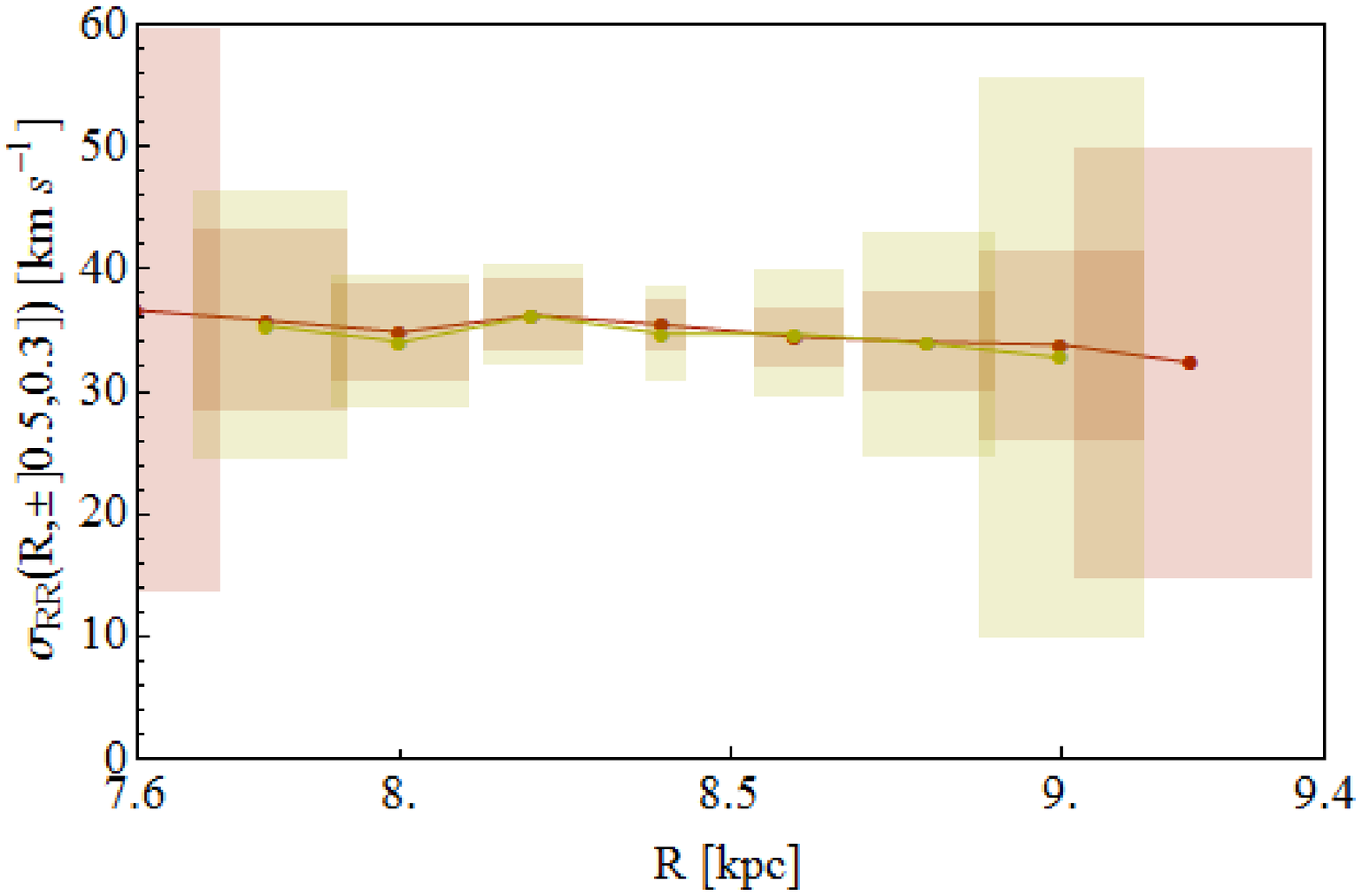}}
\caption{
Radial trend of the velocity dispersion element $\sigma _{RR} $ in the vertical interval $z \in \left]  {0.3,0.5} \right]{\rm{kpc}}$ (lighter colour) and $z \in \left]  {-0.5,-0.3} \right]{\rm{kpc}}$ (darker colour). These vertical range intervals were plotted together in order to illustrate the symmetry above and below the disk plane (at $z=0$).}
\label{SRR01}
\end{figure}

\begin{figure}
\resizebox{\hsize}{!}{\includegraphics{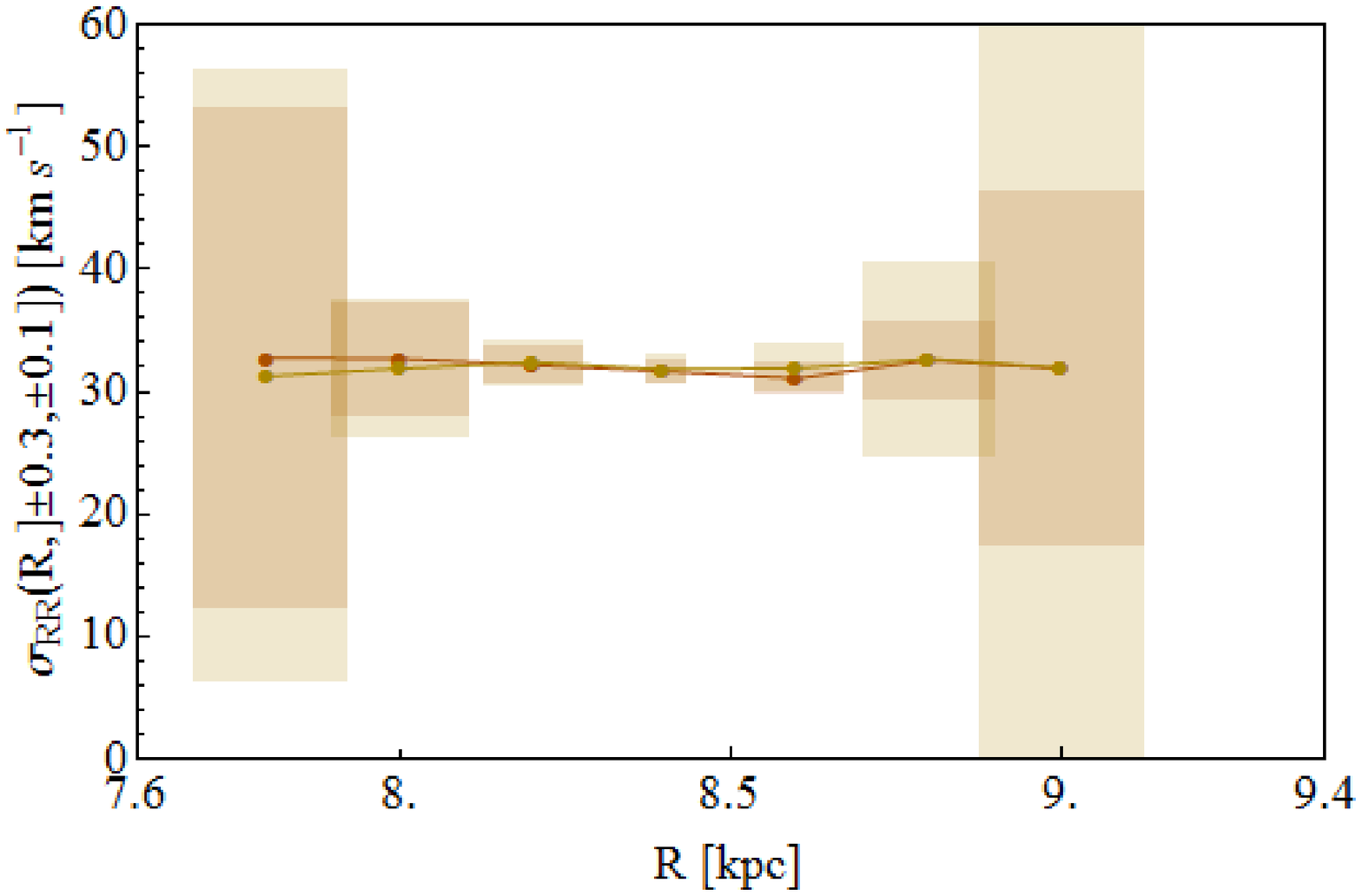}}
\caption{
Radial trend of the velocity dispersion element $\sigma _{RR} $ in the vertical interval $z \in \left]  {0.1,0.3} \right]{\rm{kpc}}$ (lighter colour) and $z \in \left]  {-0.3,-0.1} \right]{\rm{kpc}}$ (darker colour). These vertical range intervals were plotted together in order to show the  disk symmetry above and below the plane (at $z=0$).}
\label{SRR02}
\end{figure}

\begin{figure}
\resizebox{\hsize}{!}{\includegraphics{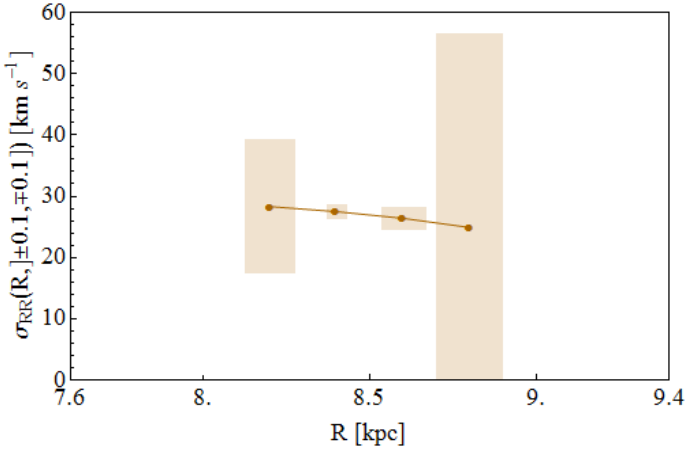}}
\caption{
Radial trend of the velocity dispersion element $\sigma _{RR} $ in the vertical interval $z \in \left] {-0.1,0.1} \right]{\rm{kpc}}$.}
\label{SRR00}
\end{figure}

 Moreover, as a general remark, we can observe how the error bars are typically smaller for the data sampled below the plane thanks to the better sampling of the RAVE data below the Galactic plane. Finally, due to the proximity of the sample to the solar location, we prefer to present a linear plot instead of a logarithmic one \citep[see, e.g.,][]{1989AJ.....97..139L,1981A&A....95..105V}. 
 The resulting trends for all the elements of the velocity dispersion tensor for the thin disk are shown in Figure \ref{Allplot}.

\begin{table*}
\caption{Trend of the velocity dispersion element $\sigma_{RR}$ expressed in $[{\rm{\text{km s}}}^{ - 1}] $. The radial distance bins (rows in leftmost column) and vertical direction bins (column headers) are in [kpc]. Empty slots indicate regions where not enough stars were sampled.}
 \centerline {\begin{tabular}{|c|r|r|r|r|r|r|}
 \hline
  $\sigma_{RR}(R,z)$   & \textbf{$<$-0.6} & \textbf{]-0.5,-0.3]} &   \textbf{]-0.3,-0.1]}    & \textbf{]-0.1,0.1]}  & \textbf{]0.1,0.3]} & \textbf{]0.3,0.5]}  \\
 \hline 
      \textbf{7.6}        &                  & $36.6\pm22.9$    &                &                &              &                  \\
      \textbf{7.8}        &  $45.6\pm18.5$   & $35.8\pm 7.3$    & $32.7\pm20.4$  &                & $31.2\pm25.0$ &  $35.3\pm10.9$    \\
      \textbf{8.0}        &  $43.6\pm13.3$   & $34.8\pm 4.0$	  & $32.6\pm 4.6$  &	 	            & $31.8\pm 5.6$ &	$34.0\pm 5.5$    \\
      \textbf{8.2}        &  $41.3\pm11.1$   & $36.2\pm 2.9$	  & $32.1\pm 1.5$  &	$28.3\pm10.9$ & $32.4\pm 1.8$ &	$36.2\pm 4.0$    \\
      \textbf{8.4}        &  $41.9\pm11.4$   & $35.4\pm 2.1$	  & $31.6\pm 1.0$  &	$27.4\pm 1.1$ & $31.8\pm 1.2$ &	$34.7\pm 3.8$    \\
      \textbf{8.6}        &  $41.7\pm13.8$   & $34.4\pm 2.4$	  & $31.0\pm 1.2$  &	$26.3\pm 1.8$ & $31.9\pm 1.9$ &  $34.7\pm 5.2$    \\
      \textbf{8.8}        &  $40.6\pm31.6$   & $34.0\pm 4.0$	  & $32.5\pm 3.1$  &	$24.8\pm31.6$ & $32.6\pm 7.9$ &	$33.9\pm 9.2$    \\
      \textbf{9.0}        &  $40.9\pm35.3$   & $33.8\pm 7.7$	  & $31.8\pm14.5$  &	              & $31.9\pm37.7$ &	$32.7\pm22.9$    \\
      \textbf{9.2}        &               	 & $32.3\pm17.6$    &                &                &              &                  \\
\hline
\end{tabular}}
\label{tabsrr}
\end{table*}

\begin{table*}
\caption{Trend of the velocity dispersion element $\sigma_{R \phi}$ expressed in $[{\rm{\text{km s}}}^{ - 1}] $. The radial distance bins (rows in leftmost column) and vertical direction bins (column headers) are in [kpc]. Empty slots indicate regions where not enough stars were sampled.}
 \centerline {\begin{tabular}{|c|r|r|r|r|r|}
 \hline
  $\sigma_{R \phi}(R,z)$  & \textbf{]-0.5,-0.3]} &   \textbf{]-0.3,-0.1]}    & \textbf{]-0.1,0.1]}  & \textbf{]0.1,0.3]} & \textbf{]0.3,0.5]}  \\
 \hline 
\textbf{7.6} & $-10.6\pm22.9$ &   &   &   &   \\
\textbf{7.8} & $-9.8\pm7.3$ & $-10.2\pm20.4$ &   & $-10.1\pm24.3$ &  $-9.7\pm10.8$ \\
\textbf{8.0} & $-10.5\pm4.0$ & $-10.2\pm4.6$ &   & $-10.0\pm5.6$ & $-9.9\pm5.5$ \\
\textbf{8.2} & $-9.6\pm2.9$ & $-10.9\pm1.6$ & $-10.2\pm10.8$ & $-10.5\pm1.9$ &  $-9.6\pm4.0$ \\
\textbf{8.4} & $-10.3\pm2.2$ & $-8.9\pm1.0$ & $-9.8\pm1.2$ & $-9.3\pm1.2$ & $-9.1\pm3.9$ \\
\textbf{8.6} & $-8.8\pm2.4$ & $-9.4\pm1.2$ & $-9.6\pm1.9$ & $-8.4\pm2.0$ & $-9.4\pm5.3$ \\
\textbf{8.8} & $-8.8\pm4.0$ & $-9.7\pm3.2$ & $-9.7\pm31.6$ & $-10.0\pm8.0$ & $-8.4\pm9.2$ \\
\textbf{9.0} & $-8.7\pm7.8$ & $-9.6\pm14.6$ &   & $-9.8\pm37.8$ & $-8.3\pm22.9$ \\
\textbf{9.2} & $-9.6\pm17.7$ &   &   &   &   \\
\hline
\end{tabular}}
\label{tabsrf}
\end{table*}

\begin{table*}
\caption{Trend of the velocity dispersion element $\sigma_{R z}$ expressed in $[{\rm{\text{km s}}}^{ - 1}] $. The radial distance bins (rows in leftmost column) and vertical direction bins (column headers) are in [kpc]. Empty slots indicate regions where not enough stars were sampled.}
 \centerline {\begin{tabular}{|c|r|r|r|r|r|}
 \hline
  $\sigma_{R z}(R,z)$  & \textbf{]-0.5,-0.3]} &   \textbf{]-0.3,-0.1]}    & \textbf{]-0.1,0.1]}  & \textbf{]0.1,0.3]} & \textbf{]0.3,0.5]}  \\
 \hline 
\textbf{7.6} &$-22.7\pm22.9$ &   &   &   &   \\
\textbf{7.8} &$-17.9\pm7.3$ & $-14.7\pm20.4$ &   & $11.7\pm24.3$ & $17.8\pm10.8$ \\
\textbf{8.0} &$-17.8\pm4.0$ & $-9.1\pm4.6$ &   & $11.0\pm5.6$ & $16.2\pm5.5$ \\
\textbf{8.2} &$-13.0\pm2.9$ & $-7.2\pm1.6$ & $-4.0\pm10.8$ & $8.3\pm1.9$ & $12.9\pm4.0$ \\
\textbf{8.4} &$-7.8\pm2.2$ & $-4.0\pm1.0$ & $-3.2\pm2.2$ & $-2.7\pm1.2$ & $9.8\pm3.9$ \\
\textbf{8.6} &$-6.1\pm2.4$ & $-2.0\pm1.2$ & $-4.5\pm2.9$ & $-4.8\pm2.0$ & $2.4\pm5.3$ \\
\textbf{8.8} &$2.8\pm4.0$ & $5.8\pm3.2$ & $-4.8\pm31.6$ & $-9.9\pm8.0$ & $-4.7\pm9.2$ \\
\textbf{9.0} &$8.1\pm7.8$ & $9.6\pm14.6$ &   & $-12.5\pm37.8$ & $-6.4\pm22.9$ \\
\textbf{9.2} &$15.4\pm17.7$ &   &   &   &   \\
\hline
\end{tabular}}
\label{tabsrz}
\end{table*}

\begin{table*}
\caption{Trend of the velocity dispersion element $\sigma_{\phi \phi}$ expressed in $[{\rm{\text{km s}}}^{ - 1}] $. The radial distance bins (rows in leftmost column) and vertical direction bins (column headers) are in [kpc]. Empty slots indicate regions where not enough stars were sampled.}
 \centerline {\begin{tabular}{|c|r|r|r|r|r|}
 \hline
  $\sigma_{\phi \phi}(R,z)$  & \textbf{]-0.5,-0.3]} &   \textbf{]-0.3,-0.1]}    & \textbf{]-0.1,0.1]}  & \textbf{]0.1,0.3]} & \textbf{]0.3,0.5]}  \\
 \hline 
\textbf{7.6} &$28.3\pm22.9$ &   &   &   &   \\
\textbf{7.8} &$27.2\pm7.3$ & $24.3\pm20.4$ &   & $24.1\pm24.3$ & $30.3\pm10.8$ \\
\textbf{8.0} &$26.8\pm4.0$ & $26.1\pm4.6$ &   & $24.2\pm5.6$ & $27.0\pm5.5$ \\
\textbf{8.2} &$24.8\pm2.9$ & $23.0\pm1.6$ & $20.3\pm10.8$ & $21.4\pm1.9$ & $26.5\pm4.0$ \\
\textbf{8.4} &$23.8\pm2.2$ & $21.7\pm1.0$ & $20.8\pm1.2$ & $20.8\pm1.2$ & $23.7\pm3.9$ \\
\textbf{8.6} &$24.8\pm2.4$ & $21.2\pm1.2$ & $19.6\pm1.9$ & $21.1\pm2.0$ & $24.2\pm5.3$ \\
\textbf{8.8} &$24.4\pm4.0$ & $21.8\pm3.2$ & $18.1\pm31.6$ & $20.0\pm8.0$ & $23.0\pm9.2$ \\
\textbf{9.0} &$24.7\pm7.8$ & $20.5\pm14.6$ &   & $17.3\pm37.8$ & $22.9\pm22.9$ \\
\textbf{9.2} &$24.2\pm17.7$ &   &   &   &   \\
\hline
\end{tabular}}
\label{tabsff}
\end{table*}

\begin{table*}
\caption{Trend of the velocity dispersion element $\sigma_{\phi z}$ expressed in $[{\rm{\text{km s}}}^{ - 1}] $. The radial distance bins (rows in leftmost column) and vertical direction bins (column headers) are in [kpc]. Empty slots indicate regions where not enough stars were sampled.}
 \centerline {\begin{tabular}{|c|r|r|r|r|r|}
 \hline
  $\sigma_{\phi z}(R,z)$  & \textbf{]-0.5,-0.3]} &   \textbf{]-0.3,-0.1]}    & \textbf{]-0.1,0.1]}  & \textbf{]0.1,0.3]} & \textbf{]0.3,0.5]}  \\
 \hline 
\textbf{7.6} &$5.1\pm22.9$ &   &   &   &   \\
\textbf{7.8} &$4.8\pm7.3$ & $3.3\pm20.4$ &   & $5.2\pm24.3$ & $4.9\pm10.8$ \\
\textbf{8.0} &$2.9\pm4.0$ & $2.1\pm4.6$ &   & $4.5\pm5.6$ & $7.3\pm5.5$ \\
\textbf{8.2} &$0.6\pm2.9$ & $4.3\pm1.6$ & $3.3\pm10.8$ & $3.4\pm1.9$ & $8.5\pm4.0$ \\
\textbf{8.4} &$-1.6\pm2.2$ & $3.7\pm1.0$ & $3.0\pm1.2$ & $4.6\pm1.2$ & $9.6\pm3.9$ \\
\textbf{8.6} &$-3.0\pm2.4$ & $5.8\pm1.2$ & $5.4\pm1.9$ & $6.0\pm2.0$ & $9.6\pm5.3$ \\
\textbf{8.8} &$-3.6\pm4.0$ & $5.3\pm3.2$ & $3.1\pm31.6$ & $5.6\pm8.0$ & $10.8\pm9.2$ \\
\textbf{9.0} &$-4.9\pm7.8$ & $4.8\pm14.6$ &   & $6.5\pm37.8$ & $10.5\pm22.9$ \\
\textbf{9.2} &$-4.7\pm17.7$ &   &   &   &   \\
\hline
\end{tabular}}
\label{tabsfz}
\end{table*}

\begin{table*}
\caption{Trend of the velocity dispersion element $\sigma_{z z}$ expressed in $[{\rm{\text{km s}}}^{ - 1}] $. The radial distance bins (rows in leftmost column) and vertical direction bins (column headers) are in [kpc]. Empty slots indicate regions where not enough stars were sampled.}
 \centerline {\begin{tabular}{|c|r|r|r|r|r|}
 \hline
  $\sigma_{z z}(R,z)$  & \textbf{]-0.5,-0.3]} &   \textbf{]-0.3,-0.1]}    & \textbf{]-0.1,0.1]}  & \textbf{]0.1,0.3]} & \textbf{]0.3,0.5]}  \\
 \hline 
\textbf{7.6} &$26.2\pm22.9$ &   &   &   &   \\
\textbf{7.8} &$25.2\pm7.3$ & $22.1\pm20.4$ &   & $22.1\pm24.3$ & $25.5\pm10.8$ \\
\textbf{8.0} &$23.1\pm4.0$ & $20.1\pm4.6$ &   & $20.8\pm5.6$ & $24.5\pm5.5$ \\
\textbf{8.2} &$20.0\pm2.9$ & $18.7\pm2.6$ & $16.7\pm10.8$ & $19.4\pm1.9$ & $20.3\pm4.0$ \\
\textbf{8.4} &$19.6\pm2.2$ & $18.0\pm2.0$ & $16.3\pm2.2$ & $17.3\pm1.2$ & $19.1\pm3.9$ \\
\textbf{8.6} &$20.0\pm2.4$ & $17.5\pm2.2$ & $14.7\pm2.9$ & $17.3\pm2.0$ & $18.6\pm5.3$ \\
\textbf{8.8} &$18.6\pm4.0$ & $17.4\pm3.2$ & $11.9\pm31.6$ & $15.7\pm8.0$ & $18.0\pm9.2$ \\
\textbf{9.0} &$16.3\pm7.8$ & $15.2\pm14.6$ &   & $14.9\pm37.8$ & $17.0\pm22.9$ \\
\textbf{9.2} &$14.6\pm17.7$ &   &   &   &   \\
\hline
\end{tabular}}
\label{tabszz}
\end{table*}

The trend for $\sigma _{zz} $  is presented in Table \ref{tabszz}. Since the original works of \citet{1982A&A...110...61V} \citep[see also][]{1991A&A...247...45A}, it is commonly assumed that, for a self-gravitating thin disk at very small $z$, a single stellar population is vertically isothermal or only slightly non-isothermal \citep[e.g.,][]{1991ApJ...368...79A,2012arXiv1202.2819B} and has a scale height nearly independent of the radius. Under this assumption and by considering an exponential dependence of the radial light profile \citep{1970ApJ...160..811F} and a constant mass-to-light ratio, the radial dependence of the vertical dispersion can be approximated as $\sigma _{zz}^2 \left( {R,0} \right) = \sigma _{zz}^2 \left( {R_ \odot  ,0} \right)e^{ - \frac{{R - R_ \odot  }}{{h_R }}} $. A different approach is based on the local constancy of Toomre's  stability parameter \citep{1964ApJ...139.1217T} which in epicyclical approximation leads to a radial dependence of the form $\sigma _{RR}^2 \left( {R,0} \right) = \sigma _{RR}^2 \left( {R_ \odot  ,0} \right)R^2 e^{ - \frac{{R - R_ \odot  }}{{h_R }}} $  \citep[e.g.,][]{1991ApJ...368...79A}. Nevertheless, to first order for the solar radial distance $R - R_ \odot  $  both the relations result in a linear fit to the plot of Fig. \ref{SRR00} \citep[see also][]{1988AJ.....95..463N}. 

A numerical solution for the velocity dispersion profile of the midplane at $z = 0$ of the MW has been found by \citet{1992MNRAS.256..166C}. Their Eqn. (22) is however dependent on the gradients of $\sigma _{zz} $ on the meridional plane or on the second derivative of the dispersion tensor. Nevertheless, our profiles can be used as a boundary condition for that type of analytical study.

\begin{figure}
\resizebox{\hsize}{!}{\includegraphics{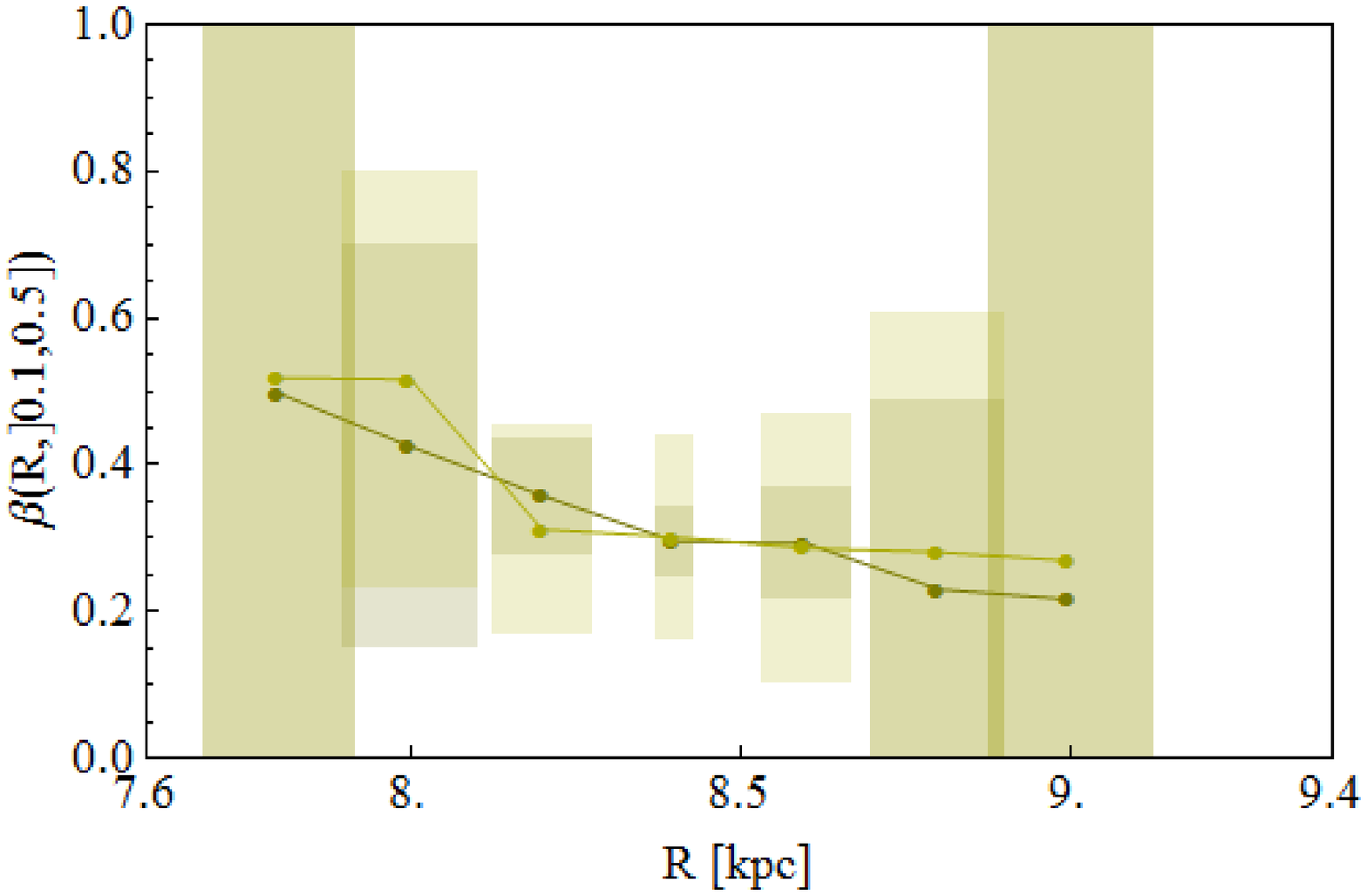}}
\caption{Trend of the anisotropy parameter $\beta$ for the vertical distance range $z \in \left] {0.3,0.5} \right]{\rm{kpc}}$ (lighter colour) and $z \in \left] { 0.1,0.3} \right]{\rm{kpc}}$ (darker colour).}
\label{beta01}
\end{figure}

\begin{figure}
\resizebox{\hsize}{!}{\includegraphics{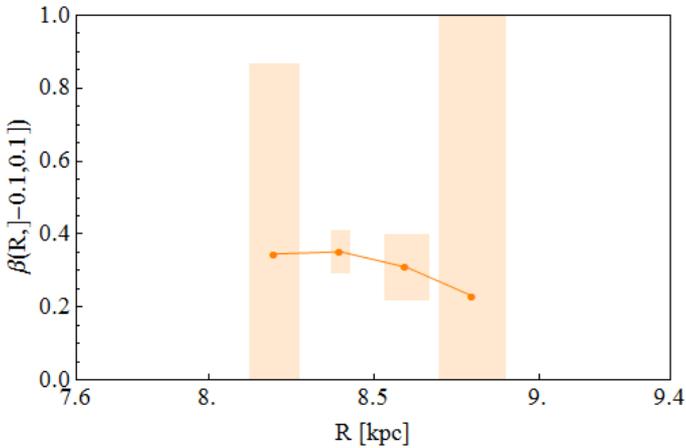}}
\caption{Trend of the anisotropy parameter $\beta$ for $z \in \left] {-0.1,0.1} \right]{\rm{kpc}}$.}
\label{beta02}
\end{figure}

\begin{figure}
\resizebox{\hsize}{!}{\includegraphics{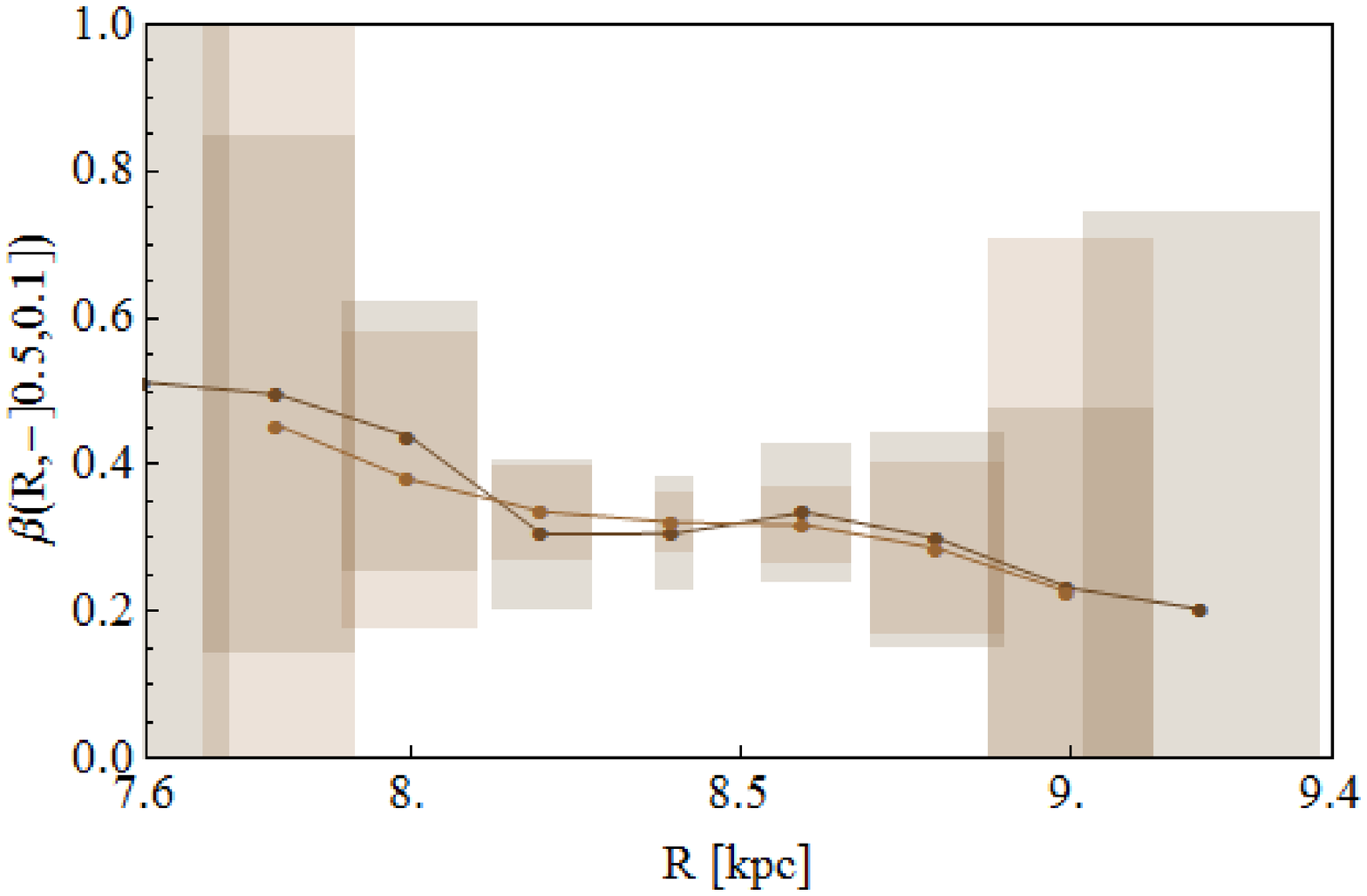}}
\caption{Trend of the anisotropy parameter $\beta$ for the vertical distance range $z \in \left] { -0.3,-0.1} \right]{\rm{kpc}}$ (lighter colour) and $z \in \left] { -0.5,-0.3} \right]{\rm{kpc}}$ (darker colour).}
\label{beta03}
\end{figure}

From the Tables \ref{tabsrr} and \ref{tabszz} we can compute directly the anisotropy parameter, which we define for the axisymmetric system as $\beta \left( {R,z} \right) \equiv \frac{{\sigma _{zz}^2 }}{{\sigma _{RR}^2 }}$.  This parameter is related to the first approach mentioned above. A plot of $\beta$ is presented in Figures \ref{beta01}, \ref{beta02} and \ref{beta03}.

\begin{figure}
\resizebox{\hsize}{!}{\includegraphics{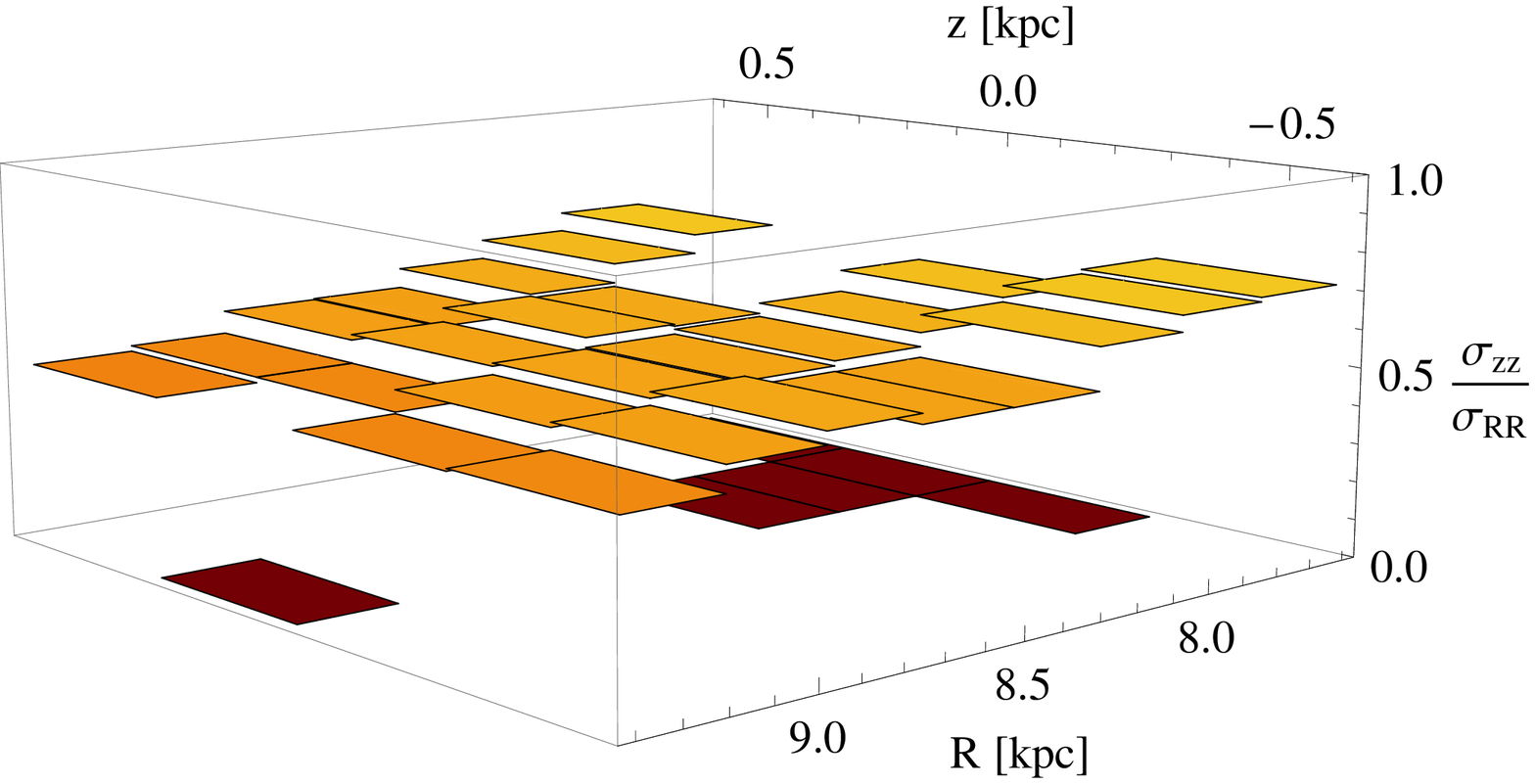}}
\caption{
Axis ratio $\sqrt \beta$ of the velocity ellipsoid in the meridional plane.}
\label{AxisRatio01}
\end{figure}

In Figure \ref{AxisRatio01} we see the trend of the axis ratio of the velocity ellipsoid \citep[e.g.,][]{1989ARA&A..27..555G} along the meridional plane. In our approach we are dealing with a composite stellar population. Thus we do not necessarily expect to match the theoretical  predictions that nevertheless remain based on several assumptions. The ratio$\frac{{\sigma _{zz} \left( {R,z} \right)}}{{\sigma _{RR} \left( {R,z} \right)}} \ne 1$
 is shown in Figure \ref{AxisRatio01}. In agreement with our initial hypotheses, once $f_{\text{thin}}  = f_{\text{thin}} \left( {E,L_z ,I_3 } \right)$ is assumed for an axisymmetric and stationary thin disk with angular momentum $L_z $ and the energy $E$, then this trend has to be interpreted as observational evidence of the dependence of $f_{\text{thin}} $ on a third integral of motion $I_3 $ and \textit{not }on a deviation from the kurtosis values expected from the Schwarzschild distribution (see Paper I for further details).
 As can been seen, in the plane $\frac{{\sigma _{zz} \left( {R,0} \right)}}{{\sigma _{RR} \left( {R,0} \right)}} \in \left[ {0.4,0.5} \right]$ and outside the plane we are close to constancy within the error bars. Nevertheless, a small trend can be observed in Fig. \ref{AxisRatio01} that can be interpreted as a suggestion of a small decrease of this ratio.

Finally in Table \ref{tabszz} the trend of $\sigma _{zz}  = \sigma _{zz} \left( {R,z} \right)$ is reported. As is evident, the thin disk distribution of the mixture of main sequence stars seems to be remarkably isothermal \citep[see, also][]{2012arXiv1202.2819B}. This is an apparent effect due to the mixture of the different populations acting to hide the velocity dispersion-age relation. The study of this relation is beyond the scope of this paper but can be easily achieved without a statistical procedure to disentangle the thin and thick disk component in order to consider a more relevant statistical example (or vice versa: age and chemistry can also be taken into consideration to improve the disentangling technique, see Sect. \ref{conclusion}). 
 
The coupling between vertical and radial direction is shown with the off diagonal term $\sigma _{Rz}  = \sigma _{Rz} \left( {R,z} \right)$ in Table \ref{tabsrz}. This represents the tilt of the velocity ellipsoid in the radial direction in cylindrical coordinates. By symmetry we expect and we confirm that on the plane this tilt is approximately zero; $\sigma _{Rz} \left( {R,0} \right) \cong 0$. The tilt angle here has the remarkable meaning to represent a projection of the third integral of motion $I_3 $
 on the surface section of the meridional plane. Using  $\Delta  = \frac{1}{2}\left( {\arctan \left( {\frac{{2\sigma _{Rz}^2 }}{{\sigma _{RR}^2  - \sigma _{zz}^2 }}} \right)} \right)$, we can obtain (from Table \ref{tabsrr} and \ref{tabszz}) a  smoothly growing trend for the tilt as we get to higher $\left| z \right|$, even though the trend is not perfectly zero at our Sun's position $z=z_{\text{Sun}}$. Nevertheless, within the errors, the values are in agreement with the independent determination presented in \citet{2006A&A...451..125V} and  the work of \citet{1991ApJ...368...79A} \citep[or with the independent, compatible formulation by][]{2009A&A...500..781B}.

\begin{figure}
\resizebox{\hsize}{!}{\includegraphics{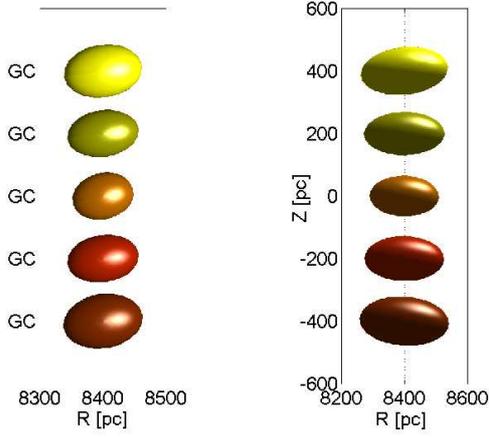}}
\caption{
Velocity ellipsoids in the solar cylinder as deduced from the velocity dispersions of Fig. \ref{Allplot} from two different points of view. On the left, we see the radial-azimuthal tilt of the velocity ellipsoids with respect to the direction of the Galactic centre (GC). The colours of the velocity ellipsoids refer to the height above or below the Galactic plane as indicated in the right panel (as well as in the vertical sequence of panels in Fig.\ref{Allplot}). The right panel shows the vertical tilt of the velocity ellipsoids above and below the plane in the solar cylinder.
}
\label{Ellipsoids02}
\end{figure}

Tables \ref{tabsrf} and \ref{tabsff}  report the values for $\sigma _{\phi \phi }  = \sigma _{\phi \phi } \left( {R,z} \right)$
 and $\sigma _{R\phi }  = \sigma _{R\phi } \left( {R,z} \right)$
 respectively. The first of these terms, $\sigma _{\varphi \varphi }  = ({\overline {v_\varphi   - \bar v_\varphi  } })^{1/2} $ represents the second central moments of the azimuthal direction distribution. Despite the bias influencing the $\bar v_\phi  $ determination, a smooth trend is still visible, which suggests an increasing velocity dispersion toward the Galactic centre as well as in the direction of increasing $\left| z \right|$
 with remarkable symmetry above and below the plane. The second term, $\sigma _{R\phi } $
is null only in the case of axisymmetry, although in the literature we found considerable misuse of this assumption in the theoretical models. Nevertheless this assumption does not hold in the case of our Galaxy as seen in Table \ref{tabsrf} and a greater effort should be made in order to relax the axisymmetry assumption in favour of a more permissive rotation symmetry about the $z$ axis for the thin disk DF, e.g.,  $f_{\text{thin}} \left( \phi  \right) = f_{\text{thin}} \left( {\phi  + \pi } \right)$. Actually, no analytical DF relaxing this approximation is known or applied. This term is strictly related to the phenomenon of the vertex deviation, i.e. the angular tilt of the velocity ellipsoid with the radial azimuthal direction by an angle $\psi $ so that $\tan \left( {2\psi } \right) = \frac{{2\sigma _{R\phi }^2 }}{{\sigma _{RR}^2  - \sigma _{\phi \phi }^2 }}$. The problem is strictly related to the problem of the streaming motions.  Its  impact on the velocity ellipsoid in the solar neighbourhood has been emphasized by several authors \citep[e.g.,][]{2003A&A...398..141S, 1998AJ....115.2384D, 1999ApJ...524L..35D}. Streaming motions can be recognized as local moving groups. \citet{2007MNRAS.380.1348S} show the presence of local moving groups complicates the age-velocity dispersion relation. As pointed out by \citet{1998AJ....115.2384D} and \citet{2004MNRAS.350..627D}, they affect the vertex deviation. Up to now, streaming motions have been studied only in the solar vicinity to get a local map of the velocity space \citep{1986RMxAA..13....9B, 1982ApJ...255..217H, 1972A&A....18...97M, 1970IAUS...38..423W, 1998AJ....115.2384D}. Nevertheless, it is not possible to perform a check with the Hipparcos data for the solar neighbourhood because the vast majority of the RAVE stars are outside of the range the Hipparcos parallaxes as well as the studies on the local vertex deviation. Finally the afore-mentioned problems with the first moment of the data distribution along the azimuthal direction is expected to influence the second central moments in this direction. In the left panel of Fig. \ref{Ellipsoids02} we show a graphical representation of the angular tilt of the velocity ellipsoid with the radial azimuthal direction. The RAVE data seem to support the idea that this tilt results as an effect of the streaming motions, showing a larger azimuthal tilt for stars closer to the plane. 

Finally Table \ref{tabsfz} shows the values derived for $\sigma _{\phi z}  = \sigma _{\phi z} \left( {R,z} \right)$. They are considerably smaller than the previous terms or null within the error bars. $\sigma _{\phi z}  = 0$ for a single stellar population tells us that the azimuthal and vertical velocities are approximately decoupled, or equivalently that the short axis of the velocity ellipsoid remains aligned with the cylindrical coordinate system (as expected at $z = 0$). The same results seem to hold for the selected sample of dwarf stars in the thin disk.

\section{Conclusions}\label{conclusion}		
\begin{figure}
\resizebox{\hsize}{!}{\includegraphics{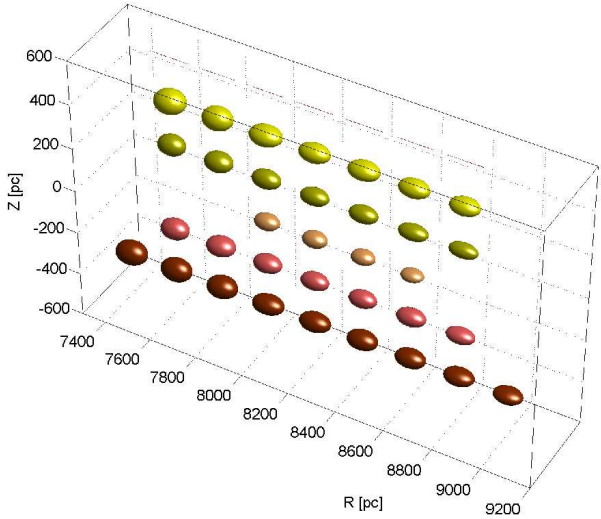}}
\caption{
Simple representation of the results on the velocity dispersion trend. The velocity dispersion tensor in each spatial point $(R,z)$ can be represented with an ellipsoidal surface defined by $\sigma _{ij}^{ - 2} v_i v_j  = 1$ (by summing over repeated indices). Here we plotted the velocity ellipsoids with their axes tabulated in the previous Tables \ref{tabsrr}, \ref{tabszz}, \ref{tabsff} and their tilt listed in Tables \ref{tabsrf}, \ref{tabsrz} and \ref{tabsfz}. The colour gradient ranges from light colours above the plane to darker colours below the plane.}
\label{Ellipsoids01}
\end{figure}

The subject of this paper is the derivation the first two moments of the velocity distribution for the thin disk component based on stellar spectra for stars in the solar neighbourhood observed by RAVE. This was realized by disentangling the thin and thick disk; the thick disk has been the subject of our first paper (Pasetto et al. 2012, submitted) and in the present paper we exclusively focus on the thin disk.
Our main goal is the investigation of the solar motion relative to the LSR and the thin disk velocity dispersion trend. We derived two of the solar motion components as 
\[
\left\{ {v_{U, \odot } ,v_{W, \odot } } \right\} = \left\{ {10.9 \pm 1.0,7.2 \pm 1.3} \right\}\;{\rm{km}}\;{\rm{s}}^{ - 1} 
\]
All the results on the thin disk velocity ellipsoid can be summarized in a single figure, Fig.  \ref{Ellipsoids01}. Here we see the radial and vertical extensions that we have been able to map with the RAVE data as well as the ellipsoid axis ratios and their tilts. All the values come from the Tables \ref{tabsrr} to \ref{tabszz}.

There are several points that need a more extensive investigation. We just mention a few here:
\begin{itemize}
	\item A methodology to obtain constraints on the disk potential and the radial velocity sample. A classical approach to this problem is based on a pre-assumed form for an analytical distribution function based on three isolating integrals of motion $f_{\text{thin}}  = f_{\text{thin}} \left( {I_1 ,I_2 ,I_3 } \right)$
 from which by a projection procedure on the phase space, one can deduce analytically observable quantities directly linked to the potential. Unfortunately, these approaches, as the one adopted in Paper I, are subject to assumptions (axisymmetry, stationarity) that fail already in the solar neighbourhood: the most obvious case is perhaps the vertex deviation but see also recent work by \citet[e.g.,][]{2011AJ....142...51L} and \citet{2011ApJ...738...27B}. Hence it would be useful to have an analytical model able to relax these assumptions in order to fully exploit the possibilities of the RAVE survey.
 \item The relation between stellar evolution, chemistry and the kinematics (the velocity dispersion) has been neglected in this paper. The reason is that in order to obtain a reasonable amount of stars and a reasonable precision in the dispersion tensor determination by mapping of the meridional plane, we should work with a limited subsample of stars closer to us. 
 Nevertheless, it is in principle possible to further narrow down the thin disk component with selection criteria based on chemical arguments. For example, \citet[][]{2009A&A...497..563N} determined the trend of abundance ratios as a function of $\left[ {\text{Fe}/\text{H}} \right]$ from 451 high-resolution spectra of F, G, and K main-sequence stars in the solar neighbourhood confirming a dichotomy in $\left[ {\alpha /\text{Fe}} \right]$ found already by \citet[e.g.,][]{2006MNRAS.367.1329R} or \citet{2004AN....325....3F}. \citet{2011MNRAS.412.1203N} investigated the possibility of disentangling thin and thick disk stars using a combination of  $\left[ {\text{Fe}/\text{H}} \right]$, $\left[ {\alpha /\text{Fe}} \right]$ and the heavy element Eu for a  sample of 306 stars. Models mixing kinematics and chemistry are currently in development \citep[e.g.,][]{2009MNRAS.396..203S}. Their results require further investigations with respect to the role of the radial migration of stars \citep[e.g.,][]{2011A&A...534A..75B, 2011ApJ...735L..46B, 2012MNRAS.420..913B} or gas \citep[e.g.,][]{2011A&A...531A..72S} and the presence of metallicity gradients. Nevertheless, the RAVE catalogue is surely suitable for this kind of investigation since more and more data are being obtained and an extension of this disentangling technique to include giant stars and chemical constraints is worthy of investigation.
 \item Although the average values for $\bar v_R $, $\bar v_\phi$ and $\bar v_z $
 present a smooth trend, we are not yet able to compute their derivatives $\partial _R \bar v_i $ or $\partial _z \bar v_i $, which remain targets of primary importance of forthcoming surveys such as Gaia \citep[e.g.,][]{2012A&A...543A.100R}. These elements are of fundamental importance in order to investigate the shape of the thin disk distribution function in order to clarify the importance of the third integral of motion over the deviation from the expected kurtosis value for a Schwarschild DF.
\end{itemize}
For the time being, however, the direct comparison of this observational work can be done only with other observational works. Finally a few cautionary remarks on the limit of our results. As already commented in the introduction, the extrapolation of the solar neighbourhood kinematic description to the whole Galactic disk relies on the validity of the Jeans theorem in the integral description of the thin disk distribution function. This is subject to the validity of the stationary condition that holds in a broad sense for the group of stars sampled by the RAVE survey. Nevertheless, we recall that several processes may cause small violations of this assumption in the solar neighbourhood, for instance:
\begin{itemize}
	\item Spiral structures and the bar redistribute angular momentum across the disk and heat it up \citep[e.g.,][]{1985ApJ...292...79C} by causing small deviations from the steady-state situation and potentially are indeed a main source influencing the second order moment of the $f_{\text{thin}} $ here studied. \citet{2010ApJ...722..112M} show an example of mixing by combined effects of bar and spirals even if the effect that they registered seems to be short-lived.
	\item The Galaxy accretes angular momentum from the remnants of external dwarf galaxies that are merging with it and the accretion is expected to lead to significant reorientation of the spin axis \citep[e.g.,][]{1986MNRAS.218..743B,2002MNRAS.336..785S} by potentially influencing the trend of the first moment of $f_{\text{thin}} $. 
\end{itemize}
Other processes that are not naturally considered in the stationary description include, e.g., the influence of the Outer Lindblad Resonance of the Galactic bar in the solar neighborhood  \citep[e.g.,][]{1999ApJ...524L..35D} or the constant stripping of small stellar streams from satellite galaxies \citep[e.g.,][]{1999MNRAS.307..495H}. These processes are expected to influence the data in the RAVE catalogue when also the giant stars are taken into consideration.

\begin{acknowledgements}
S.P. wants to thank B. Fuchs, A. Just and J. Binney for comments on the technique and results of the paper and I. Minchev for careful reading of the manuscript.
Numerical computations have been partially performed with supercomputers at the John von Neumann - Institut f$\ddot{u}$r Computing (NIC) - Germany  (NIC-project number 2979). We acknowledge partial funding from Sonderforschungsbereich SFB 881 ``The Milky Way System'' (subprojects A5 and A6) of the German Research Foundation (DFG).
Funding for RAVE has been provided by: the Australian Astronomical
Observatory; the Leibniz-Institut fuer Astrophysik Potsdam (AIP); the
Australian National University; the Australian Research Council; the
French National Research Agency; the German Research Foundation (SPP 1177 and SFB 881); 
the European Research Council (ERC-StG 240271 Galactica); the Istituto Nazionale di 
Astrofisica at Padova; The Johns Hopkins University; the National Science Foundation of 
the USA (AST-0908326); the W. M. Keck foundation; the Macquarie University; the 
Netherlands Research School for Astronomy; the Natural Sciences and Engineering Research 
Council of Canada; the Slovenian Research Agency; the Swiss National Science Foundation; 
the Science \& Technology Facilities Council of the UK; Opticon; Strasbourg Observatory; 
and the Universities of Groningen, Heidelberg and Sydney. The RAVE web site is at 
http://www.rave-survey.org.
\end{acknowledgements}
\bibliographystyle{aa}
\bibliography{BiblioArt}

\end{document}